\documentclass[acmtog]{acmart}

\usepackage{booktabs} 

\usepackage[ruled]{algorithm2e} 

\SetAlFnt{\small}
\SetAlCapFnt{\small}
\SetAlCapNameFnt{\small}
\SetAlCapHSkip{0pt}

\setcopyright{rightsretained}
\acmJournal{TOG}
\acmYear{2020}\acmVolume{39}\acmNumber{4}\acmArticle{112}\acmMonth{7} \acmDOI{10.1145/3386569.3392447}

\usepackage[T1]{fontenc} 
\usepackage{epsfig}
\usepackage{graphicx}
\usepackage{amsmath}
\usepackage{amssymb}
\usepackage{color}
\usepackage{stmaryrd}
\usepackage{float}
\usepackage{caption}
\usepackage{subcaption}
\usepackage{booktabs}
\usepackage{url}
\usepackage{wrapfig}

\def\textrightarrow{ -> }%

\usepackage{mathtools}
\usepackage{algorithmicx}

\algnewcommand{\LineComment}[1]{\State{\emph{// #1}}}

\newcommand{\algspace}[0]{\vspace{.05in}}

\algrenewcommand\alglinenumber[1]{
	{\sf\footnotesize\addfontfeatures{Colour=888888,Numbers=Monospaced}#1}}
\usepackage[ruled]{algorithm2e} 

\SetAlFnt{\footnotesize}
\SetAlCapFnt{\footnotesize}
\SetAlCapNameFnt{\footnotesize}
\SetAlCapHSkip{0pt}
\IncMargin{-\parindent}
\usepackage{wrapfig}

\newlength{\commentWidth}
\SetCommentSty{mycommfont}

\newcommand{\newT}[1]{{#1}}

\newcommand{\CH}[1]{{#1}}

\newcommand{\xR}{\mathbb{R}}
\newcommand{\tin}{\!\in\!}

\newcommand{\atRow}[2]{{\left[{#1}\right]_{{#2}\cdot}} }
\newcommand{\pwise}[0]{T}
\newcommand{\pwiseMat}[0]{P}
\newcommand{\fmap}[0]{C}
\newcommand{\redBasis}[0]{\Psi}
\newcommand{\redBasisSize}{k}

\newcommand{\vertSet}{\mathcal{V}}
\newcommand{\mV}{\vertSet}
\newcommand{\faceSet}{\mathcal{F}}
\newcommand{\edgeSet}{\mathcal{E}}
\newcommand{\Src}[0]{M_1}
\newcommand{\Tar}[0]{M_2}
\newcommand{\vertexCoords}{X}
\newcommand{\mX}{\vertexCoords}
\newcommand{\nVerts}{n}
\newcommand{\area}{a}

\newcommand{\nLands}{m}
\newcommand{\landSet}{\mathcal{S}}
\newcommand{\landSetAss}{\landSet^+}
\newcommand{\landSetUnProc}{\landSet^-}
\newcommand{\landAss}{l^+}
\newcommand{\landUnProc}{l^-}
\newcommand{\adjSet}{\mathcal{A}}
\newcommand{\origSet}{\mathcal{O}}
\newcommand{\lndEps}{d_{\epsilon}}
\newcommand{\lndM}{\mu}
\newcommand{\dAdj}{d_{\text{adj}}}
\newcommand{\descDist}{\mathcal{W}}
\newcommand{\epsWks}{\epsilon_{\text{wks}}}

\newcommand{\chrom}{c}
\newcommand{\chromChild}{\bar{c}}

\newcommand{\chromTarOf}[1]{[#1]}
\newcommand{\matchSize}{\tilde{m}}
\newcommand{\geneBank}{\mathcal{G}}
\newcommand{\pcross}{$p_{cross}$}
\newcommand{\pmutG}{$p_{grow}$}
\newcommand{\pmutS}{$p_{shrink}$}
\newcommand{\pmutFMG}{$p_{FMguid}$}

\newcommand{\ElaplComm}{\mathcal{E}^{\Delta}}
\newcommand{\Elands}{\mathcal{E}^m}
\newcommand{\Erevers}{\mathcal{E}^{\text{rev}}}
\newcommand{\Eelastic}{\mathcal{E}^{\text{elstc}}}
\newcommand{\Efit}{\mathcal{E}^{\text{fit}}}
\newcommand{\Efmap}{\mathcal{E}^{\text{map}}}
\newcommand{\Emembrane}{\mathcal{E}^{\text{mem}}}
\newcommand{\Ebending}{\mathcal{E}^{\text{bnd}}}

\begin{document}
	
	\title[ENIGMA: Evolutionary Non-Isometric Geometry Matching]{ENIGMA: Evolutionary Non-Isometric Geometry MAtching}
	
	\author{Michal Edelstein}
	\affiliation{%
		\institution{Technion - Israel Institute of Technology, Israel}}
	\email{smichale@cs.technion.ac.il}
	
	\author{Danielle Ezuz}
	\affiliation{%
		\institution{Technion - Israel Institute of Technology, Israel}}
	\email{danielle.ezuz@gmail.com}
	
	\author{Mirela Ben-Chen} 
	\affiliation{%
		\institution{Technion - Israel Institute of Technology, Israel}}
	\email{mirela@cs.technion.ac.il}

	

	\begin{abstract}

		In this paper we propose a \emph{fully automatic} method for shape correspondence 
		that is widely applicable, and especially effective for non isometric shapes and shapes of different topology. We observe that fully-automatic shape correspondence can be decomposed as a hybrid discrete/continuous optimization problem, and we find the best \emph{sparse} landmark correspondence, whose sparse-to-dense extension minimizes a \emph{local metric distortion}. To tackle the \emph{combinatorial} task of landmark correspondence we use an evolutionary \emph{genetic algorithm}, where the local distortion of the sparse-to-dense extension is used as the objective function.
		We design novel \emph{geometrically} guided genetic operators, which, when combined with our objective, are highly effective for non isometric shape matching.	 Our method outperforms state of the art methods for automatic shape correspondence both quantitatively and qualitatively on challenging datasets.
	\end{abstract}


	
	\maketitle

	\section{Introduction}
	Shape correspondence is a fundamental task in shape 
	analysis:
	given two shapes, the goal is to compute a semantic correspondence between 
	points on them.
	Shape correspondence is required when two shapes are analyzed jointly, which is 
	common in many applications such as texture and deformation 
	transfer~\cite{sumner2004deformation}, statistical shape 
	analysis~\cite{munsell2008evaluating} and shape 
	classification~\cite{ezuz2017gwcnn}, to mention just a few examples.
	
    The difficulty of the shape matching problem depends on the class of \emph{deformations} that can be applied to one shape to align it with the second. For example, if only \emph{rigid} transformations are allowed it is easier to find a correspondence than if non-rigid deformations are also possible, since the number of degrees of freedom is small and the space of allowed transformations is easy to parameterize. Similarly, if only \emph{isometric} deformations are allowed, the matching is easier than if non-isometry is possible, since then there is a clear criteria of the quality of the map, namely the preservation of geodesic distances. The hardest case is when the two shapes belong to the same semantic class, but are not necessarily isometric. In this case, the correspondence algorithm should achieve two goals: (1) put in correspondence semantically meaningful points on both shapes, and (2) reduce the \emph{local} metric distortion.
	
	Hence, the non-isometric shape correspondence problem is often considered 
	as a \emph{two step} process. First, the \emph{global semantics} of the 
	matching is given by a sparse set of \emph{corresponding landmarks} of salient points on both shapes. If this set is informative enough, then 
	the full shapes can be matched by extending the landmark correspondence to a 
	full map from the source to the target in a \emph{consistent} and \emph{smooth} 
	way. The first problem is \emph{combinatorial}, requiring the computation of a 
	permutation of a subset of the landmarks, whereas the second problem is 
	\emph{continuous}, requiring the definition and computation of local 
	differential properties of the map. Whereas the second problem has been tackled 
	by multiple 
	methods~\cite{aigerman2016hyperbolic,mandad2017variance,RHM,ezuz2019elastic} 
	which yield excellent results for non-isometric shapes, methods that address 
	the sparse landmark correspondence 
	problem~\cite{kezurer2015tight,Maron2016point,DSppDym,SahilliogluGenetic} have 
	so far been limited either to the nearly isometric case, or to a very small set 
	of landmarks.
	
		\begin{figure}[t]
		\includegraphics[width=\linewidth]{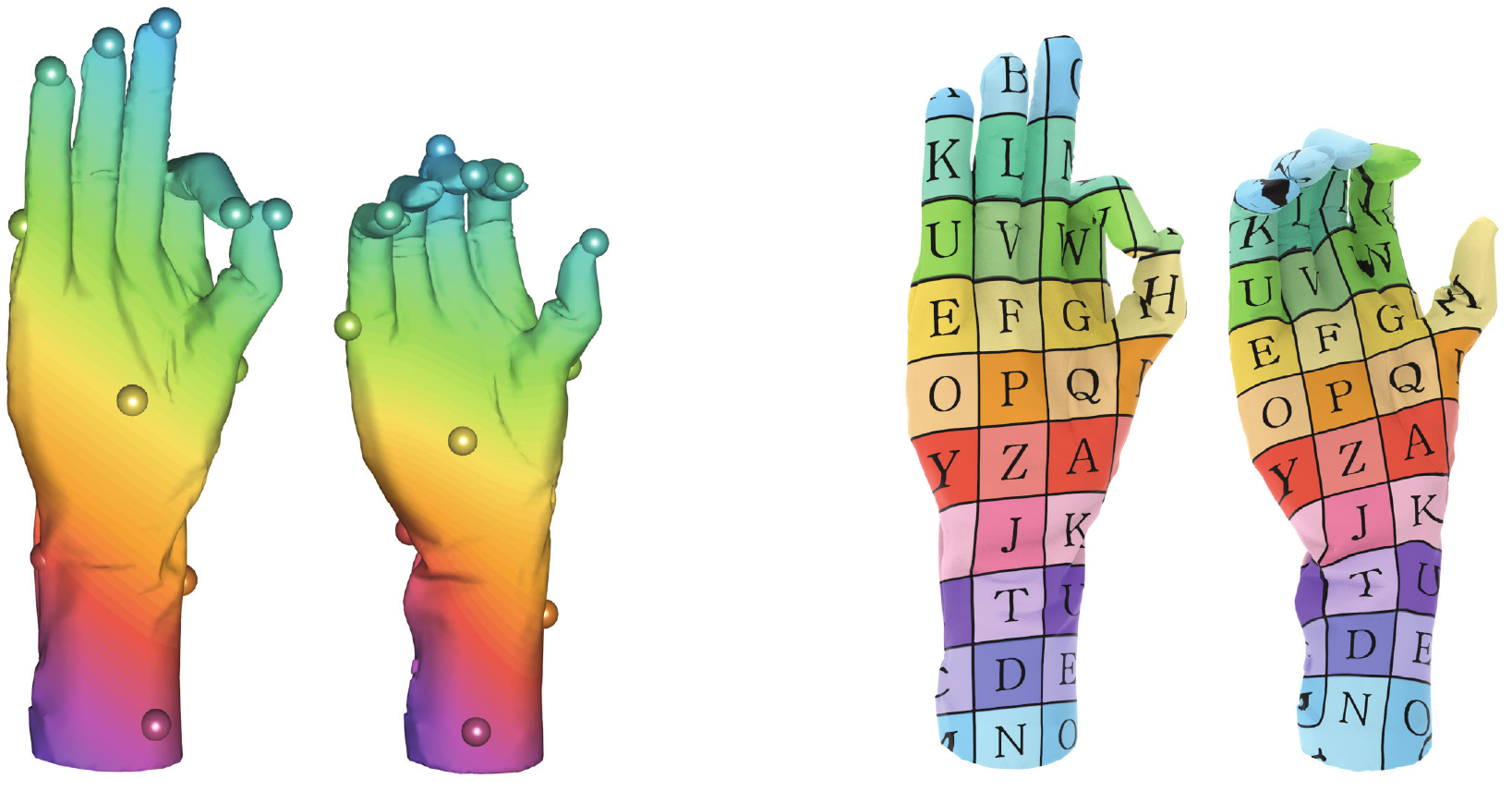}
		\centering
		\caption{A map between shapes of different genus obtained by our approach. 
			(left) Output landmark correspondence and functional map, visualized 
			using color transfer. (right) Final pointwise map visualized using texture transfer. }
		\label{fig:teaser1}
	\end{figure}
	
We propose to leverage the efficient algorithms for solving the second problem to generate a framework for solving the first. Specifically, we suggest a combinatorial optimization for matching a sparse set of landmarks, such that the best obtainable local distortion of the corresponding sparse-to-dense extension is minimized. As the optimization tool, we propose to use a \emph{genetic algorithm}, as these have been used for combinatorial optimization 	for a few decades~\cite{holland1992genetic}, and are quite general in the type 	of objectives they can optimize. Despite their success in other fields, though, to the best of our knowledge, their 	use in shape analysis has been limited so far to isometric 
	matching~\cite{SahilliogluGenetic}. 
	
	Using a 
	genetic algorithm allows us to optimize a challenging objective function, that is 
	given both in terms of the landmark permutation and the differential properties 
	of the extended map computed from these landmarks. We use a non-linear 
	non-convex objective, given by the \emph{elastic energy} of the deformation of 
	the shapes, an approach that has been recently used successfully for 
	non-isometric matching~\cite{ezuz2019elastic} when the landmarks are known. Furthermore, we apply the \emph{functional map} framework~\cite{ovsjanikov2012functional} to allow for an efficient computation of the elastic energy. Finally, paramount to the success of a genetic algorithm are the genetic operators, that \emph{combine} two sparse correspondences to generate a new one, and \emph{mutate} an existing correspondence. We design novel \emph{geometric} genetic operators that are guaranteed to yield new valid correspondences. We show that our algorithm yields a landmark 
	correspondence that, when extended to a functional map and a full 
	vertex-to-point map, outperforms existing state-of-the-art techniques for 
	automatic shape correspondence, both quantitatively and qualitatively.

	\subsection{Related Work}
	As the literature on shape correspondence is vast, we discuss here only methods 
	which are directly relevant to our approach. For a more detailed survey on 
	shape correspondence we refer the reader to the excellent 
	reviews~\cite{van2011survey,tam2013registration}.
	
	\paragraph*{Fully automatic shape correspondence.}
	Many fully automatic methods, like ours, compute a \emph{sparse} correspondence 
	between the shapes, to decrease the number of degrees of freedom and possible 
	solutions. A sparse-to-dense method can then be used in a post processing step 
	to obtain a dense map. For example, one of the first of such methods was proposed by Zhang et al.~\shortcite{Zhang2008deformation}, who used a search tree traversal technique to optimize a deformation energy of sparse landmark correspondence.
	Later, Kezurer et al.~\shortcite{kezurer2015tight} formulated the shape 
	correspondence problem as a Quadratic Assignment Matching problem, and 
	suggested a convex semi-definite programming (SDP) relaxation to solve it 
	efficiently. While the convex relaxation was essential, their method is still only 
	suitable for small sets (of the same size) of corresponding landmarks. Dym et 
	al.~\shortcite{DSppDym} suggested to combine doubly stochastic and spectral 
	relaxations to optimize the Quadratic Assignment Matching problem, which is not 
	as tight as the SDP relaxation, but much more efficient.
	Maron et al.~\shortcite{Maron2016point} suggested a convex relaxation to 
	optimize a term that relates pointwise and functional maps, which promotes 
	isometries by constraining the functional map to be orthogonal.

	Other methods for the automatic computation of a dense map include Blended 
	Intrinsic Maps (BIM) by Kim et al.~\shortcite{kim2011blended}, who optimized for the most 
	isometric combination of conformal maps. Their method works well for relatively 
	similar shapes and generates locally smooth results, yet is restricted to genus-$0$ surfaces.
	Vestner et al.~\shortcite{vestner2017product} suggested a multi scale approach that 
	is not restricted to isometries, but requires shapes with the same number of 
	vertices and generates a bijective vertex to vertex correspondence.

	A different approach to tackle the correspondence problem is to compute a 
	\emph{fuzzy} map~\cite{ovsjanikov2012functional, solomon2016entropic}. The 
	first approach puts in correspondence functions instead of points, whereas the 
	second is applied to probability distributions. These generalizations allow much 
	more general types of correspondences, e.g. between shapes of different genus, 
	however, they also require an additional pointwise map extraction step.
	The functional map approach was used and extended by many following methods, 
	for example Nogneng et al.~\shortcite{nogneng2017informative} introduced a pointwise 
	multiplication preservation term, Huang et al.~\shortcite{huang2017adjoint} used the 
	adjoint operators of functional maps, and Ren et al.~\shortcite{Ren2018Continuous} 
	recently suggested to incorporate an orientation preserving term and a 
	pointwise extraction method that promotes bijectivity and continuity, that can 
	be used for non isometric matching as well.

	Our method differs from most existing methods by the quality measure that we 
	optimize. Specifically, we optimize for the landmark correspondence by 
	measuring the elastic energy of the \emph{dense} correspondence implied by 
	these landmarks. As we show in the results section, our approach outperforms 
	existing automatic state-of-the-art techniques on challenging datasets.
	
	\paragraph*{Semi-automatic shape correspondence.}
	Many shape correspondence methods use a non-trivial initialization, e.g., a 
	sparse landmark correspondence, to warm start the optimization of a dense 
	correspondence. 
	Panozzo et al.~\shortcite{panozzo2013weighted} extended a given landmark correspondence by computing the surface barycentric coordinates of a source point with respect to the landmarks on the source shape, and matching that to target points which have similar coordinates with respect to the target landmarks. The landmarks were chosen interactively, thus promoting intuitive user 
	control. More recently, Gehre et al.~\shortcite{gehre2018interactive}, used curve 
	constraints and functional maps, for computing correspondences in an 
	interactive setting. Given landmark correspondences or an extrinsic alignment, 
	Mandad et al.~\shortcite{mandad2017variance} used a soft correspondence, where each 
	source point is matched to a target point with a certain 
	probability, and minimized the variance of this distribution. 
	Parameterization based methods map the two shapes to a single, simple domain, 
	such that the given landmark correspondence is preserved, and the composition 
	of these maps yields the final 
	result~\cite{aigerman2015seamless,aigerman2015orbifold,aigerman2016hyperbolic}.
	Since minimizing the distortion of the maps to the common domain does not 
	guarantee minimal distortion of the final map, recently Ezuz et al.~\shortcite{RHM} 
	directly optimized the Dirichlet energy and the reversibility of the forward 
	and backward maps, which led to results with low conformal distortion.
	Finally, many automatic methods, for example, all the functional map based 
	approaches, can use landmarks as an auxiliary input to improve the results in 
	highly non isometric settings.
	
	The output of our method is a sparse correspondence and a functional map, which 
	can be used as input to semi-automatic methods to generate a dense map. Thus, 
	our approach is complementary to semi-automatic methods. Specifically, we 
	use a recent, publically available method~\cite{RHM} to extract a dense vertex-to-point map from the landmarks and 
	functional map that we compute with the genetic algorithm.
	
	\paragraph*{\CH{Learning based Methods.}}	

	\CH{	
	Since the correspondence problem can be difficult to model analytically, many machine learning and neural networks based methods, have been suggested.
	Litman et al.~\shortcite{litman2014learning} learn shape descriptors based on their spectral properties. Wei et al.~\shortcite{wei2016dense} find the correspondence between humans' depth maps rendered from multiple viewpoints by computing per-pixel feature descriptors. \cite{huang2017learning} learn a local descriptor from multiple images of the shapes, taken from different angles and scales. Other methods use local isotropic or anisotropic filters in intrinsic convolution layers~\cite{masci2015geodesic,boscaini2016learning,monti2017geometric}. Lim et al.~\shortcite{lim2018simple} presented SpiralNet, performing convolution by a spiral operation for enumerating the information from neighboring vertices. Poulenard and Ovsjanikov \shortcite{poulenard2018multi} define a multi-directional convolution.
	 Another approach, proposed by Litany et al.~\shortcite{litany2017deep}, is a network architecture based on functional maps, which was additionally used for \emph{unsupervised} learning schemes~\cite{halimi2019unsupervised,roufosse2019unsupervised}.
	}

	\paragraph*{Genetic algorithms.}
	
	Genetic algorithms were initially inspired by the process of evolution and 
	natural selection~\cite{holland1992genetic}.
	In the last few decades they have been used in many domains, such as: protein folding 
	simulations~\cite{unger1993genetic}, clustering~\cite{maulik2000genetic} and 
	image segmentation~\cite{bhanu1995adaptive}, to mention just a few. In the context of graph and shape 
	matching, genetic algorithms were used for registration of depth 
	images~\cite{chow2004surface,silva2005precision}, 2D shape recognition in 
	images~\cite{ozcan1997partial}, rigid registration of 3-D curves and 
	surfaces~\cite{yamany1999new}, and inexact graph 
	matching~\cite{cross1997inexact,auwatanamongkol2007inexact}.
	
	More recently, Sahillio\u{g}lu~\shortcite{SahilliogluGenetic} suggested a 
	genetic algorithm for isometric shape matching. However, their approach is 
	very different from ours. First, their objective was the preservation of 
	the pairwise distances between the \emph{sparse} landmarks, which is only appropriate for isometric shapes. We, on the the other hand, use the energy of a \emph{dense} correspondence, which is the output of a sparse-to-dense algorithm, as our objective. This allows us to match correctly shapes which have large non-isometric deformations.   
Furthermore, we define novel geometric genetic operators, which are tailored to our problem, and lead to superior results in challenging, highly non-isometric cases. We additionally compare with other state of the art methods for automatic sparse 	correspondence~\cite{kezurer2015tight,DSppDym,SahilliogluGenetic}, and with a recent functional map based method~\cite{Ren2018Continuous} that automatically 
	computes dense maps and does not have topological restrictions. We apply the 
	same sparse-to-dense  post-processing to all methods, and 
	show that we outperform previous methods, as demonstrated by both quantitative 
	and qualitative evaluation.

	\subsection{Contribution}
	Our main contributions are:
		\begin{itemize}			
			\item A novel use of the functional framework within a genetic pipeline for designing an efficient, geometric objective function that is resilient to non-isometric deformations.
			
			\item Novel geometric genetic operators for \emph{combining} and \emph{mutating} partial sparse landmark correspondences that are guaranteed to yield valid new correspondences.
			
			\item A fully automatic pipeline for matching non-isometric shapes that achieves superior results compared to state-of-the-art automatic methods.
		\end{itemize}
	
	\section{Background}
	\subsection{Notation}
   \paragraph{Meshes.}	
	We compute a correspondence between two manifold triangle meshes, denoted by 
	$\Src$ and $\Tar$. The vertex, edge and face sets are denoted by $\vertSet_i,\ \edgeSet_i,\  	\faceSet_i$, respectively, where the subscript $i \tin \{1,2\}$ indicates the 
	corresponding mesh. We additionally denote the number of vertices $|\vertSet_i|$ by $\nVerts_i$. 	The embedding in $\xR^3$ of the mesh $M_i$ is given by $\mX_i \tin 
	\xR^{\nVerts_i \times 3}$. Given a matrix $A \tin \xR^{\nVerts \times r}$, we denote its $v$-th row by $\atRow{A}{v} \tin \xR^{1 \times r}$. For example, the embedding in $\xR^3$ of a vertex $v \tin \vertSet_i$ is given by $\atRow{\mX_i}{v} \tin \xR^{1 \times 3}$. 
	The area of a face $f \tin \faceSet_i$, and a vertex $v \tin \vertSet_i$ are denoted by $\area_f, \area_v$, respectively. The area of a vertex is defined to be a third of the sum of the areas of its adjacent faces.
	
	\paragraph{Maps.} A pointwise map that assigns a point on $\Tar$ to each \emph{vertex} of $\Src$ is denoted 
	by $\pwise_{12}:\vertSet_1 \to \Tar$. The corresponding functional map matrix that maps piecewise linear functions from $\Tar$ to $\Src$ is denoted by  
	$\pwiseMat_{12}\tin \mathbb{R}^{\nVerts_1 \times \nVerts_2}$.
	Similarly, maps in the opposite direction are denoted by swapped subscripts, e.g. 
	$\pwise_{21}$ is a pointwise map from $M_2$ to $M_1$.	
	The eigenfunctions of the Laplace-Beltrami operator of $M_i$ that correspond to the 
	smallest $\redBasisSize_i$ eigenvalues are used as a reduced 
	basis for scalar functions, and are stacked as the columns of the basis matrix 
	$\redBasis_i\tin \mathbb{R}^{\nVerts_i \times \redBasisSize_i}$. 
	A functional map matrix that maps functions  in these reduced bases from $\Tar$ to $\Src$ is denoted by $\fmap_{12}\tin 
	\mathbb{R}^{\redBasisSize_1 \times \redBasisSize_2}$.

	\subsection{Functional Maps}
	\label{sec:fmap_background}
	\subsubsection{Maps as Composition Operators.}
	Given a pointwise map 	$ \pwise_{12}$ that maps vertices on $\Src$ to points on $\Tar$, let $\mX_{12} \tin \xR^{n_1 \times 3}$ denote the embedding coordinates in $\xR^3$ of the mapped points on $\Tar$. Now, consider a vertex $v_1 \tin \vertSet_1$ such that $\pwise_{12}(v_1)$ lies on the triangle $ \{u_2, v_2, w_2\} \tin \faceSet_2$, where $u_2, v_2, w_2 \tin \vertSet_2$. We can represent the embedding coordinates of $\pwise_{12}(v_1)$ using \emph{barycentric coordinates} as $\atRow{\mX_{12}}{v_1} = \gamma_u \atRow{\mX_2}{u_2} + \gamma_v \atRow{\mX_2}{v_2} + \gamma_w \atRow{\mX_2}{w_2} \tin \xR^{1\times 3}$, where $\gamma_u, \gamma_v, \gamma_w$ are non-negative and sum to 1. Alternatively, we can write this concisely as $\atRow{\mX_{12}}{v_1} = \gamma X_2$, where $\gamma \tin \xR^{1 \times n_2}$ is a vector that has all entries zero except at the indices $u_2, v_2, w_2$, which hold the barycentric coordinates $\gamma_u, \gamma_v, \gamma_w$, respectively. 	
	We repeat this process for all the vertices of $M_1$, and build a matrix $\pwiseMat_{12} \tin \xR^{\nVerts_1 \times \nVerts_2}$, such that, for example, $\atRow{\pwiseMat_{12}}{v_1} = \gamma$. Now, it is easy to check that $\mX_{12} = \pwiseMat_{12} \mX_2$. 
	
	In the same manner	that we have applied it to the coordinate functions $\mX_2$, the matrix $\pwiseMat_{12}$ can  be used to map any \emph{piecewise linear function} from $\Tar$ to $\Src$. Specifically, given a piecewise linear function on $\Tar$, which is given by its values at the vertices $f_2 \tin \xR^{\nVerts_2 \times 1}$, the corresponding function on $\Src$ is given by $f_1 = \pwiseMat_{12} f_2 \tin \xR^{\nVerts_1 \times 1}$. The defining property of $f_1$ is that is given by \emph{composition}, specifically $f_1(v_1) = f_2(\pwise_{12}(v_1))$, where we extend $f_2$ linearly to the interior of the faces. Thus, $\pwiseMat_{12} f_2 = f_2 \circ \pwise_{12}$. 
	
	The idea of using composition operators, denoted as \emph{functional maps}, to represent maps between surfaces, was first introduced by Ovsjanikov et 	al.~\shortcite{ovsjanikov2012functional,ovsjanikov2016computing}. Note that, on triangle meshes, if $\pwise_{12}$ is given, $\pwiseMat_{12}$ is defined \emph{uniquely}, using the barycentric coordinates construction that we have described earlier. If, on the other hand, $\pwiseMat_{12}$ is given, and \emph{it has a valid sparsity structure and values}, then it also uniquely defines the map $\pwise_{12}$. This holds, since the non-zero indices in $\atRow{\pwiseMat_{12}}{v_1}$ indicate on which triangle of the target surface $\pwise_{12}(v_1)$ lies, and the non-zero values indicate the barycentric coordinates of the mapped point in that triangle.

	Working with the matrices $\pwiseMat_{12}$ instead of the pointwise maps $\pwise_{12}$ has various advantages~\cite{ovsjanikov2016computing}. One of them is that instead of working with the full basis of piecewise linear functions, one can work in a \emph{reduced basis} $\redBasis_i$ of size $\redBasisSize_i$. By conjugating $\pwiseMat_{12}$ with the reduced bases, we get a compact \emph{functional map}:
	\begin{equation}
	\fmap(\pwiseMat_{12}) = \fmap_{12} = \redBasis_1^\dagger \pwiseMat_{12} 
	\redBasis_2 \in \mathbb{R}^{k_1\times k_2}.
	\label{eq:pwise_to_fmap}
	\end{equation} 

The small sized matrix $\fmap_{12}$ is easier to optimize for than the full matrix $\pwiseMat_{12}$. Furthermore, for simplicity, the structural constraints on $\pwiseMat_{12}$ are not enforced when optimizing for $\fmap_{12}$, leading to unconstrained optimization problems with a small number of variables. Conversely, once a small matrix $\fmap_{12}$ is extracted, a full matrix $\pwiseMat_{12}$ is often reconstructed and converted to a pointwise map $\pwise_{12}$ for downstream use in applications. The reconstruction is \emph{not} unique (see, e.g., the discussion by Ezuz et al.~\shortcite{ezuz2017deblurring}), and in this paper we will use:
	\begin{equation}
	\pwiseMat(\fmap_{12}) = \pwiseMat_{12} = \redBasis_1 \fmap_{12} 
	\redBasis_2^\dagger \in \mathbb{R}^{n_1\times n_2}.
	\label{eq:fmap_to_pwise}
	\end{equation} 
Note that $\pwiseMat(\fmap_{12})$ does not, in general, have the required structure to represent a pointwise map $\pwise_{12}$. We will address this issue further in following sections.

	\subsubsection{Basis.}
	The first $\redBasisSize_i$ eigenfunctions of the Laplace-Beltrami (LB) 
	operator of $M_i$ are often used as the reduced basis $\redBasis_i$, such that smooth functions are well approximated 	using a small number of coefficients and $\fmap_{12}$ is compact.
	
	It is often valuable to use a larger basis size for the \emph{target functions}, so that mapped functions are well represented. Hence, since $\fmap_{ij}$ maps functions on $M_j$ to functions on $M_i$, $\redBasis_i$ 
	should contain more basis functions than $\redBasis_j$. Thus, we denote the number of source eigenfunctions by $k_s$, the number of target eigenfunctions 
	by $k_t$, and the functional maps in both directions $\fmap_{12},\fmap_{21}$are of size $k_t \times k_s$. Slightly abusing notations, and to avoid clutter, 
	we use $\redBasis_i$ to denote the eigenfunctions corresponding to $M_i$, in 
	both directions, namely both when $k_i = k_s$ and $k_i = k_t$, as the meaning 
	is often clear from the context. Where required, we will explicitly denote by 
	$\redBasis_{i_s}, \redBasis_{i_t}$ the eigenfunctions with dimensions $k_s, 
	k_t$, respectively.
	For example, $\fmap(\pwiseMat_{12}) = \fmap_{1_t 2_s} = \redBasis_{1_t}^\dagger \pwiseMat_{12} \redBasis_{2_s} \tin \xR^{\redBasisSize_t\times\redBasisSize_s}$, and similarly for $\pwiseMat(\fmap_{12})$.
	
	\subsubsection{Objectives.}	
	Many cost functions have been suggested for functional map computation, 
	e.g.,~\cite{ovsjanikov2012functional,nogneng2017informative,cheng2018geometry,Ren2018Continuous},
	among others. In our approach we use the following terms.

	\paragraph*{Landmark correspondence.}  Given a set $\Pi$ of pairs of 
	corresponding landmarks, $\Pi = \{(i,j) \,|\, i \in \mV_1, j \in \mV_2\}$, we 
	use the term:
	\begin{equation}
	\Elands \left( \fmap_{12}, \Pi \right) = \sum _{\left(i,j\right)\in \Pi}\| 
	 \atRow{\redBasis_1}{i} \fmap_{12} -  \atRow{\redBasis_2}{j} \| ^2.
	\label{eq:Elands}
	\end{equation}
	While some methods use landmark-based descriptors, we prefer to avoid it due to 
	possible bias towards isometry that might be inherent in the descriptors. The 
	formulation in Equation~\eqref{eq:Elands} has been used successfully by Gehre 
	et al.~\cite{gehre2018interactive} for functional map computation between 
	highly non isometric shapes, as well as in the context of pointwise map 
	recovery~\cite{ezuz2017deblurring}.

	\paragraph*{Commutativity with Laplace-Beltrami.} We use:
	\begin{equation}
	\ElaplComm \left( \fmap_{12} \right) = \| \Delta_1 \fmap_{12} - \fmap_{12} 
	\Delta_2 \|_F ^2,
	\label{eq:ElaplComm}
	\end{equation}
	where $\Delta_i$ is a diagonal matrix holding the first $k_i$ eigenvalues of the 
	Laplace-Beltrami operator of $M_i$. 
	While initially this term was derived to promote isometries, in practice it has 
	proven to be useful for highly non isometric shape matching as 
	well~\cite{gehre2018interactive}.
	
	To compute a functional map $\fmap_{12}$ from a set $\Pi$ of pairs of corresponding landmarks, we optimize the following combined objective:
	\begin{equation}
	\label{eq:fmap}
	\Efmap(\fmap_{12}, \Pi) = \alpha \, \ElaplComm(\fmap_{12}) + \beta \, \Elands( \fmap_{12}, 
	\Pi).
	\end{equation}
	\subsection{Elastic Energy}
	\label{sec:elastic}
	Elastic energies are commonly used for shape 
	deformation~\cite{botsch2006primo,sorkine2007as,HeRuWa12,HeRuSc14}, but were 
	used for shape matching as 
	well~\cite{DrLiRuSc05,windheuser2011geometrically,DEBUHAN2016783,iglesias2015shape,ezuz2019elastic}.
	In this paper we use a recent formulation that achieved state of the art 
	results for non isometric shape matching~\cite{ezuz2019elastic}. The full 
	details are described there, and we mention here only the main equations for 
	completeness. 
	
	The elastic energy is defined for a source undeformed mesh $M$, and a 
	deformed mesh with the same triangulation but different geometry $\tilde{M}$.
	It consists of two terms, a \emph{membrane} term and a \emph{bending} term. 	
	The membrane energy penalizes area distortion:
	\begin{equation}
	\Emembrane ( M, \tilde{M} ) =  \sum_{t \in \faceSet} a_t \left( 
	\frac{1}{2} \mathrm{tr} \mathcal{G}_t + \frac{1}{4} \det \mathcal{G}_t - 
	\frac{3}{4} \log \det \mathcal{G}_t - \frac{5}{4} \right) ,
	\end{equation}
	where $\area_t$ denotes the area of face $t$, and $\mathcal{G}_t$\CH{$=\hat{g}_t^{-1}g_t\in \xR^{2\times 2}$} denotes the 
	geometric distortion tensor of the face $t$. \CH{Here $\hat{g}_t,g_t$ are the discrete first fundamental forms of $M$ and $\tilde{M}$, respectively}.
	 In addition, in order to have a finite expression in case some triangles are deformed into zero area triangles, 
	the negative $\log$ function is linearly extended below a small threshold. 
	
	The bending energy, on the other hand, penalizes misalignment of curvature 
	features:
	\begin{equation}
	\Ebending ( M, \tilde{M} ) = \sum_{e \in \edgeSet} 
	\frac{(\tilde\theta_e - \theta_e)^2}{\tilde d_e} \left( \tilde 
	l_e\right) ^2\, ,
	\end{equation}
where $\theta_e, \tilde\theta_e$ denote the dihedral angle at edge $e$ in the 
	undeformed and deformed surfaces respectively; if $t,t'$ are the adjacent 
	triangles to $e$ then $\tilde d_e = \frac{1}{3}(\tilde{a}_t + \tilde{a}_{t'})$, 
	and $\tilde l_e$ is the length of $e$ in the deformed surface. To handle degeneracies, we sum only over edges where both adjacent triangles are not degenerate.

	The elastic energy is:
	\begin{equation}
	\label{eq:elastic}
	\Eelastic ( M, \tilde M ) = \mu \, \Emembrane ( M, \tilde M 
	) + \eta \, \Ebending ( M, \tilde M  ) \, ,
	\end{equation}
	where we always use $\mu = 1, \eta = 10^{-3}$.
	
	Given a pointwise map $\pwise_{12}: \Src \to \Tar$, we evaluate the induced elastic energy as follows. The undeformed mesh is given by $\Src$, and 
	the geometry of the deformed mesh is given by the embedding on $\Tar$ of the images of the vertices of $\Src$. Specifically, these are given 
	by $\pwise_{12}(\mV_1)$, or equivalently, $\pwiseMat_{12} \mX_2$, where $\mX_2$ 
	is the embedding of the vertices of $M_2$. 
	
	\subsection{Energies of Functional Maps}
	\label{sec:elasticEFmap}
	\paragraph*{Elastic energy.}
	The elastic energy can also be used to evaluate functional maps directly, by 
	setting the geometry of the deformed mesh to $\pwiseMat(\fmap_{12}) 
	\vertexCoords_2$.
	For brevity, we denote this energy by $\Eelastic \left( \fmap_{12} \right)$.
	
	\paragraph*{Reversibility energy.}
	In~\cite{ezuz2019elastic} the elastic energy was symmetrized and combined with 
	a \emph{reversibility} term that evaluates bijectivity. The reversibility term 
	requires computing both $\fmap_{12}$ and $\fmap_{21}$. 
	Here we define the reversibility energy for functional maps, similarly 
	to~\cite{eynard2016coupled}:
	\begin{equation}
	\label{eq:rev}
	\begin{split}	
	\Erevers \left( \fmap_{12}, \fmap_{21} \right) = &\| \fmap_{12} 
	\redBasis_{2}^\dagger P(\fmap_{21}) \vertexCoords_1  - \redBasis_{1}^\dagger 
	\vertexCoords_1 \|_F ^2 +\\
	&\| \fmap_{21} \redBasis_{1}^\dagger \pwiseMat(\fmap_{12}) \vertexCoords_2  - 
	\redBasis_{2}^\dagger \vertexCoords_2 \|_F ^2 \ .
	\end{split}
	\end{equation} 
	
	The reversibility energy measures the distance between vertex coordinates 
	(projected on the reduced basis), and their mapping to the other shape and 
	back. The smaller this distance is, the more bijectivity is promoted~\cite{RHM}.
	\begin{figure*}[t!]
		\centering
		\includegraphics[width=\linewidth]{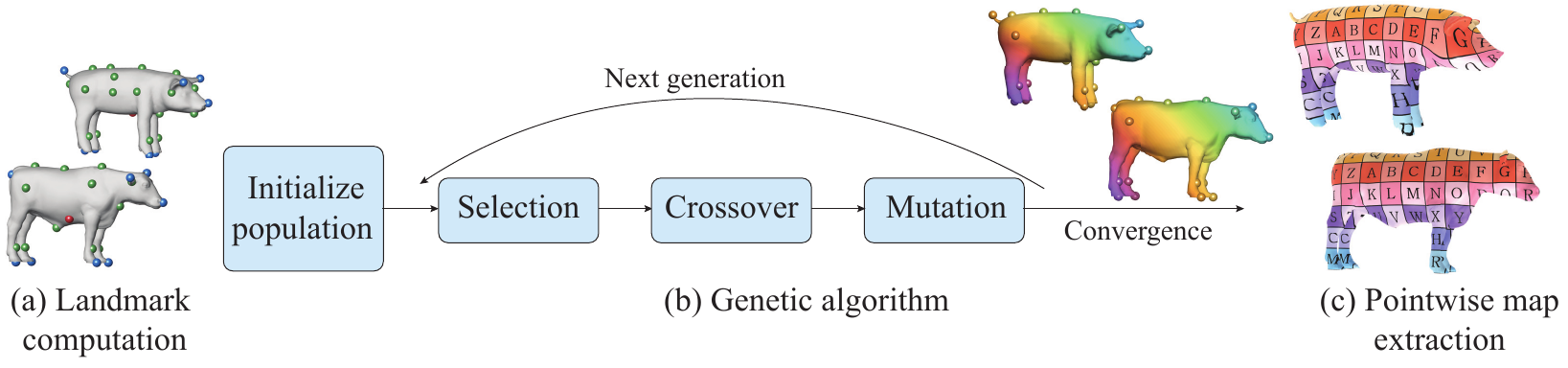}
		\caption{Our pipeline: (a) landmark computation 
			(Section~\ref{sec:landmark_comp}), (b) genetic algorithm 
			(Section~\ref{sec:genetic}), (c) sparse to dense post processing 
			using~\cite{RHM}.} 
		\label{fig:pipeline}
	\end{figure*}
	
	\section{Method}
	
	\label{sec:pipeline}
	Our goal is to automatically compute a semantic correspondence between two 
	shapes, denoted by $M_1$ and $M_2$. The shapes are the only input to our 
	method. We do not assume that the input shapes are isometric, but we do assume 
	that both shapes belong to the same semantic class, so that a semantic 
	correspondence exists.
	Our pipeline consists of three main steps, see Figure~\ref{fig:pipeline}:
	\begin{enumerate}
		\item[(a)] Compute two sets of geometrically meaningful landmarks on $M_i$, 
		denoted by $\landSet_i$ (Section~\ref{sec:landmark_comp}).
		\item[(b)] Compute a \emph{partial} sparse correspondence, i.e. a permutation $\Pi$ between subsets of the landmarks $\landSetAss_i \subseteq \landSet_i$, as well as corresponding functional 
		maps $\fmap_{ij}$, using a genetic algorithm (Section~\ref{sec:genetic}).
		\item[(c)] Generate a dense pointwise map using an existing semi automatic 
		correspondence method~\cite{RHM}.
	\end{enumerate}
	
	We use standard techniques for the first and last steps, and thus our main 
	technical contribution lies in the design of the genetic algorithm.
	
	The most challenging problem is determining the objective 
	function. 
	Unlike the isometric correspondence case, where it is known that pairwise 
	distances between landmarks should be preserved, in the general (not 
	necessarily isometric) case there is no known criterion that the \emph{landmarks} 
	should satisfy.
	However, there exist well studied \emph{differential} quality measures of local distortion, which have proved to be useful in practice for \emph{dense} non isometric correspondence.
    Hence, our approach is to \emph{find a landmark correspondence that induces the best distortion minimizing map}.  	
	The general formulation of the optimization problem we address is therefore:	
	\begin{equation}
	\begin{aligned}
	& \underset{\Pi}{\text{minimize}}
	& & \Efit(\fmap_{\text{opt}}(\Pi)) \\
	& \text{subject to}
	& & \fmap_{\text{opt}}(\Pi) = \underset{\fmap}{\text{argmin}} \,\, 
	\Efmap(\fmap, \Pi).
	\end{aligned}
	\label{eq:opt}
	\end{equation}
	Here, $\Efit$ is a non-linear, non-convex objective, measured on the extension of the landmarks to a full map, and $\Efmap$ is the objective used for computing that extension. Thus, the variable in the objective $\Efit$, i.e. the induced map given by a set of landmarks, is itself the solution of an optimization problem. Hence, two important issues come to light.
	
	First, the importance of using a genetic algorithm, that, in addition to handling combinatorial variables, can be applied to very general objectives. And second, the importance of \emph{efficiently} evaluating $\Efit$ and solving the interior optimization problem for the extended map. We achieve this efficiency by using a \emph{functional map approach}, which performs most computations in a reduced basis, and is thus significantly faster than pointwise approaches.

	In the following we discuss the details of the algorithm, first addressing the 
	landmark computation, and then the design of the genetic algorithm that we use 
	to optimize Equation~\eqref{eq:opt}. All the parameters are fixed for all the shapes, do not require tuning, and their values are given in the Appendix~\ref{sec:Appendix}.

	\section{Automatic Landmark Computation}
	\label{sec:landmark_comp}
	As a pre-processing step, we 
	normalize both meshes to have area $1$.
	We classify the landmarks into three categories, based on their 
	computation method: \emph{maxima}, \emph{minima}, and \emph{centers}.
	
	\paragraph*{Maxima and Minima.} The first two categories are the local maxima 
	and minima of the Average Geodesic Distance (AGD), that is frequently used in 
	the context of landmark computation~\cite{kim2011blended,kezurer2015tight}. 
	The AGD of a vertex $v \tin \mV$ is defined as:
	$AGD(v)=\sum_{u\in \vertSet} \area_{u}d(v,u)$
	where $\area_{u}$ is the vertex area and $d(v,u)$ is the geodesic distance 
	between $v$ and $u$. 
	To efficiently approximate the geodesic distances, we compute a high 
	dimensional embedding, as suggested by Panozzo et 
	al.~\shortcite{panozzo2013weighted}, and use the Euclidean distances in the 
	embedding space. 
	The maxima of AGD are typically located at tips of sharp features, and the 
	minima at centers of smooth areas, thus the maxima of the AGD provide the most 
	salient features.
	
	\paragraph*{Centers.}As the maxima and minima of the AGD are very sparse, we 
	add additional landmarks using the local minima of the function: 
	\begin{equation}
	f_N(v)=\sum_{k\leq N}{\frac{1}{\sqrt{\lambda_k}} \frac{\left| \redBasis_k(v) 
			\right|}{\| \redBasis_k \| _{L^\infty} } }\ \quad v \in \mV \, ,
	\label{eq:anomalies}
	\end{equation} 
	defined by Cheng et al.~\shortcite{cheng2018geometry}, who referred to these minima 
	as \emph{centers}. Here, $\redBasis_k$ and $\lambda_k$ are the $k^{\text{th}}$ 
	eigenfunction and eigenvalue of the Laplace-Beltrami operator, and $N$ is 
	the number of eigenfunctions. 
	We use the minima of Equation~\eqref{eq:anomalies} rather than, e.g., farthest 
	point sampling~\cite{kezurer2015tight}, as the centers tend to be more 
	consistent between non isometric shapes.

	\paragraph*{Filtering.} Landmarks that are too close provide no additional 
	information, and add unnecessary degrees of freedom. Hence, we filter the 
	computed landmarks so that the minimal distance between the remaining ones is 
	above a small threshold $\lndEps$. When filtering, we prioritize the landmarks 
	according their salience, namely we prefer maxima of the AGD, then minima and 
	then centers. Finally, if the set of landmarks is larger than a maximal size of 
	$\lndM$, we increase $\lndEps$ automatically to yield less landmarks.

	\paragraph*{Adjacent landmarks.} We later define genetic operators 
	that take the geometry of the shapes into consideration. To that end, we 
	define two landmarks $l_i, r_i \tin \landSet_i$ to be \emph{adjacent} if their 
	geodesic distance $d_i(l_i, r_i) < \dAdj$ or if they are neighbors in the 
	geodesic Voronoi diagram of $M_i$.
	The set of adjacent landmarks to $l_i$ is denoted by $\adjSet(l_i)$.
	
	\paragraph*{Landmark origins.} The landmark categories are additionally used in 
	the genetic algorithm. Given a landmark $l_1 \tin\landSet_1$, we denote by $\origSet(l_1) \subseteq \landSet_2$ the landmarks on $M_2$ from the same category.
	\medskip
	
	The maxima, minima and centers of each shape that remain after filtering form 
	the landmark sets $\landSet_1, \landSet_2$. The number of landmarks is denoted 
	by $\nLands_1, \nLands_2$, and is not necessarily the same for $M_1, M_2$.

	Figure~\ref{fig:pipeline}(a) shows example landmark sets, where the color indicates the landmark type: blue for maxima, red for minima and green for 
	centers.
	As expected, the landmarks do not entirely match, however there exists a 
	substantial \emph{subset} of landmarks that do match.
	At this point the landmark correspondence is not known, and it is 
	automatically computed in the next step, using the genetic algorithm.
	
	\section{Genetic Non Isometric Maps}
	\label{sec:genetic}
	
	Genetic algorithms are known to be effective for solving challenging 
	combinatorial optimization problems with many local minima. 
	In a genetic algorithm~\cite{holland1992genetic} solutions are denoted by 
	\emph{chromosomes}, which are composed of \emph{genes}. In the initialization 
	step, a collection of chromosomes, known as the \emph{initial population} is 
	created. The algorithm modifies, or \emph{evolves}, this population by 
	\emph{selecting} a random subset for modification, and then combining two 
	chromosomes to generate a new one (\emph{crossover}), and modifying 
	(\emph{mutating}) existing chromosomes. 
	The most important part of a genetic algorithm is the objective that is 
	optimized, or the \emph{fitness function}. The ultimate goal of the genetic 
	algorithm is to find a chromosome, i.e. a solution, with the best fitness 
	value. The general genetic algorithm is described in 
	Algorithm~\ref{alg:genetic}.
	
	Genetic algorithms are quite general, as they allow the fitness function to be 
	any type of function of the input chromosome. We leverage this generality to 
	define a fitness function that is itself the result of an optimization problem. 
	We further define the genes, chromosomes, crossover and mutation operators and 
	initialization and selection strategies, in a geometric manner.
	
	\begin{figure}[t!]
		\centering
		\includegraphics[width=\linewidth]{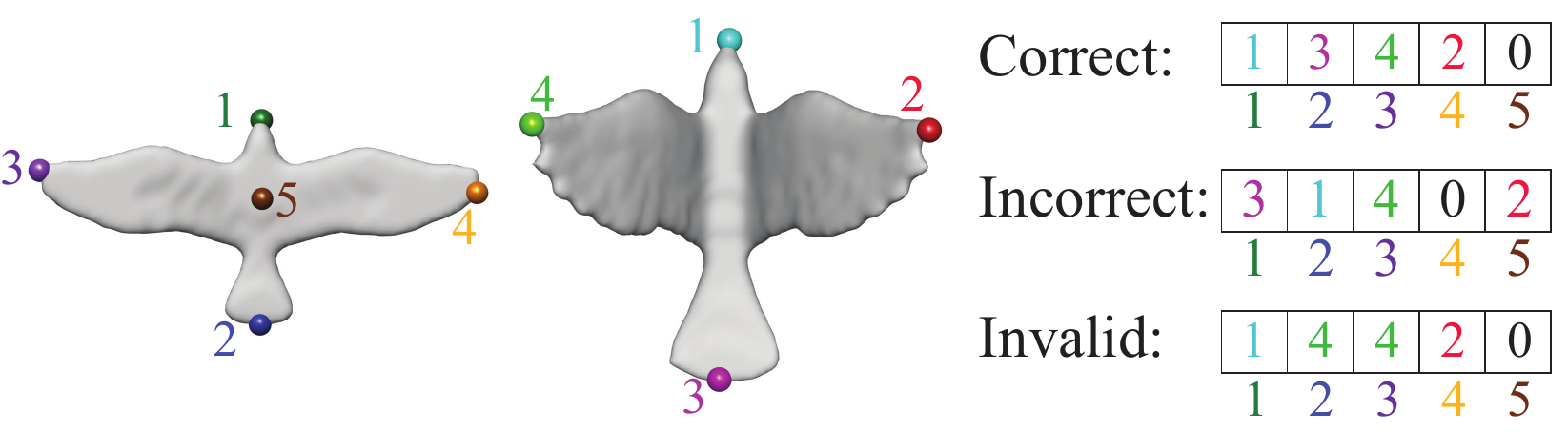}
		\caption{Illustration of chromosomes, which represent a sparse 
			correspondence. The index of an array entry corresponds a landmark of $M_1$ 
			(left), and the value corresponds to a landmark on $M_2$ (right shape). The 
			first chromosome is the desired semantic correspondence, the second is 
			valid but semantically incorrect, and the third is invalid because it is 
			not injective (landmarks 2,3 on $M_1$ correspond to the same landmark 4 on $M_2$).} 
		\label{fig:chromosome}
	\end{figure}
		\begin{algorithm}[b]
		\SetKwInOut{Input}{input}\SetKwInOut{Output}{output}
		
		\Input{$M_1, M_2, \landSet_1, \landSet_2$}
		\Output{$\Pi, \fmap_{\text{opt}}$}
		Initialize population \tcp*[r]{Section~\ref{sec:initialization}} 
		Evaluate fitness \tcp*[r]{Section~\ref{sec:fitness}} 
		\While{population did not converge}{ 
			Select individuals for breeding \tcp*[r]{Section~\ref{sec:selection}} 
			Perform crossover \tcp*[r]{Section~\ref{sec:crossover}} 
			Perform mutation \tcp*[r]{Section~\ref{sec:mutation}} 
			Evaluate offspring fitness and add to population \\
		}
		Compute output from fittest chromosome
		
		\caption{Genetic Algorithm.\vspace{-.2in}}\label{alg:genetic}
	\end{algorithm}
	
	\subsection{Genes and Chromosomes}
	\label{sec:genes}
	\paragraph*{Genes.} A \emph{gene} is given by a pair $(l_1, l_2)$ such that 
	$l_1 \tin \landSet_1$ and $l_2 \tin \landSet_2 \cup {0}$, and encodes a single 
	landmark correspondence. If $l_2 = 0$ we denote it as an \emph{empty} gene, 
	otherwise it is a \emph{non-empty} gene.
	
	\paragraph*{Adjacency Preserving Genes.} Two non-empty genes $(l_1, l_2)$ and 
	$(r_1,r_2)$ are defined to be \emph{adjacency preserving (AP) genes}, if 
	$l_i,r_i$ are adjacent landmarks on $M_i$, for $i \tin \{1,2\}$.
	
	\paragraph*{Chromosome.} A \emph{chromosome} is a collection of exactly $m_1$ 
	genes, that includes a single gene for every landmark in $\landSet_1$. A 
	chromosome is \emph{valid} if it is injective, namely each landmark on 
	$\landSet_2$ is assigned to at most a single landmark in $\landSet_1$. We 
	represent a chromosome using 
	an integer array of size $\nLands_1$, and denote it by $\chrom$, thus the gene 
	$(l_1, l_2)$ is encoded by $\chrom \chromTarOf{l_1} = l_2$ (see 
	Figure~\ref{fig:chromosome}).
	
	\paragraph*{Match.} A \emph{match} defined by a chromosome $\chrom$ is denoted 
	by $\Pi(\chrom)$ and includes all the non-empty genes in $\chrom$. The sets 
	$\landSetAss_i(\Pi) \subseteq \landSet_i$ are the landmarks that participate in 
	the genes of $\Pi$, i.e., all the landmarks that have been assigned.

	\subsection{Initial Population}
	\label{sec:initialization}
	
	There are various methods for initialization of genetic algorithms, Paul et 
	al.~\shortcite{paul2013performance} discusses and compares different initialization 
	methods of genetic algorithms for the Travelling Salesman Problem, where the 
	chromosome definition is similar to ours.
	Based on their comparison and the properties of our problem, we use a 
	\emph{gene bank}~\cite{wei2007parallel}, i.e. for each source landmark we 
	compute a subset of target landmarks that are a potential match.
	
	To compare between landmarks on the two shapes we use a descriptor based 
	distance, the Wave Kernel Signature (WKS)~\cite{aubry2011wave}. 
	While this choice can induce some isometric bias, as we generate 
	\emph{multiple} matches for each source landmark, this bias does not affect our 
	results. Let $\descDist(l_1,l_2)$ denote the normalized WKS distance between 
	two landmarks $l_1 \tin \landSet_1, l_2 \tin \landSet_2$, such that the 
	distance range is $[0,1]$, and is normalized separately for each landmark.

	\paragraph*{Gene bank.} The \emph{gene bank} of a landmark $l_1 \tin 
	\landSet_1$, denoted by $\geneBank(l_1)$ is the set of genes that match $l_1$ 
	to a landmark $l_2 \tin \landSet_2$ which is close to it in WKS distance, and 
	is of the same origin. Specifically:
	\begin{equation}
	\begin{aligned}
	\geneBank(l_1) = \, \{ (l_1, l_2) \,\, | & \,\, l_2 \in \origSet(l_1) \,\, 
	\text{\small{and}} \\
	&\,\, \descDist(l_1,l_2) < \epsWks \,\, \text{\small{and}} \,\, 
	\descDist(l_2,l_1) < \epsWks \},
	\end{aligned}
	\end{equation}
	
	The gene bank defines an initial set of possibly similar landmarks. Since it is used for the initialization, the set does not need to be accurate and the WKS distance provides a reasonable initialization even for non-isometric shapes.
	
	\paragraph*{Prominent landmark.} A landmark $l_1 \tin \landSet_1$ is denoted as 
	a \emph{prominent landmark} if its gene bank $\geneBank(l_1)$ is not empty, and 
	has at most $4$ genes. This indicates that our confidence in this landmark is 
	relatively high, and we will prefer to start the chromosome building process 
	from such landmarks.
	
	Finally, to add more genes to an existing chromosome, we will need the 
	following definition.

	\begin{wrapfigure}{r}{0.08\textwidth}
		\vspace{-.035\textheight}
		\begin{center}
			\hspace*{-20pt}
		\includegraphics[width=1.25\linewidth]{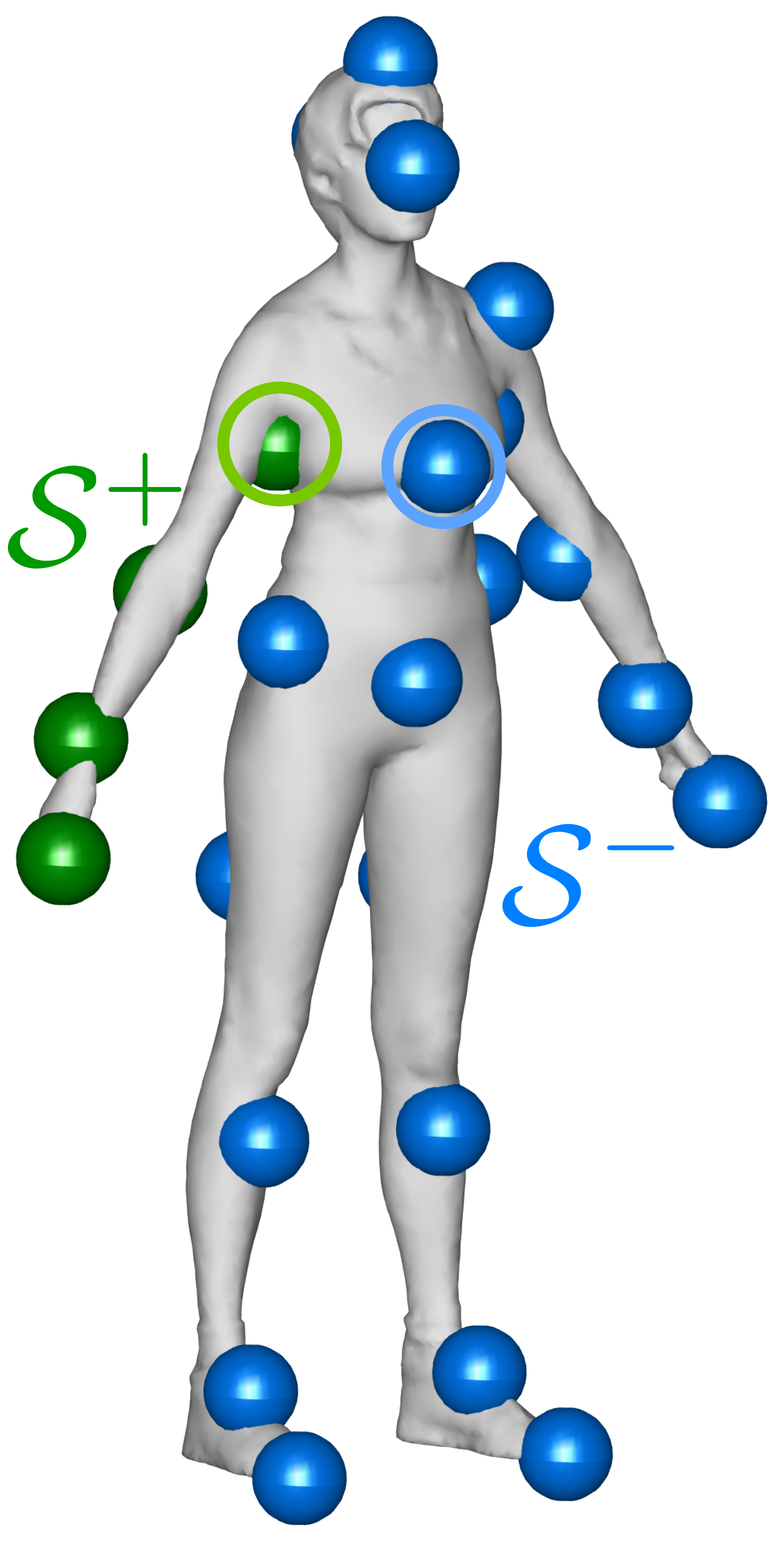}
	\end{center}
	\end{wrapfigure}

	\paragraph*{Closest matched/unmatched pair.} Given two disjoint subsets 
	$\landSetAss_1, \landSetUnProc_1 \!\subseteq\! \landSet_1$, we define the 
	\emph{closest matched/unmatched pair} as the closest adjacent pair, where each 
	landmark belongs to a different subset.
	In the inset figure, $\landSetAss_1$ is displayed in green and $\landSetUnProc_1$ is displayed in blue. The \emph{closest matched/unmatched pair} $[\landAss_1, \landUnProc_1]$ is circled, $\landAss_1$ in green and $\landUnProc_1$ in blue. Explicitly:
	
	\begin{equation}
	[\landAss_1, \landUnProc_1] = \underset{l_1 \in \landSetAss_1, r_1 \in 
		\landSetUnProc_1}{\text{argmin}} \,\, d_1(l_1, r_1) \quad \text{s.t.} \quad r_1 
	\in \adjSet(l_1).
	\end{equation}

	\subsubsection{Chromosome construction} 
	
\begin{figure*}
		\centering
		\includegraphics[width=0.95\linewidth]{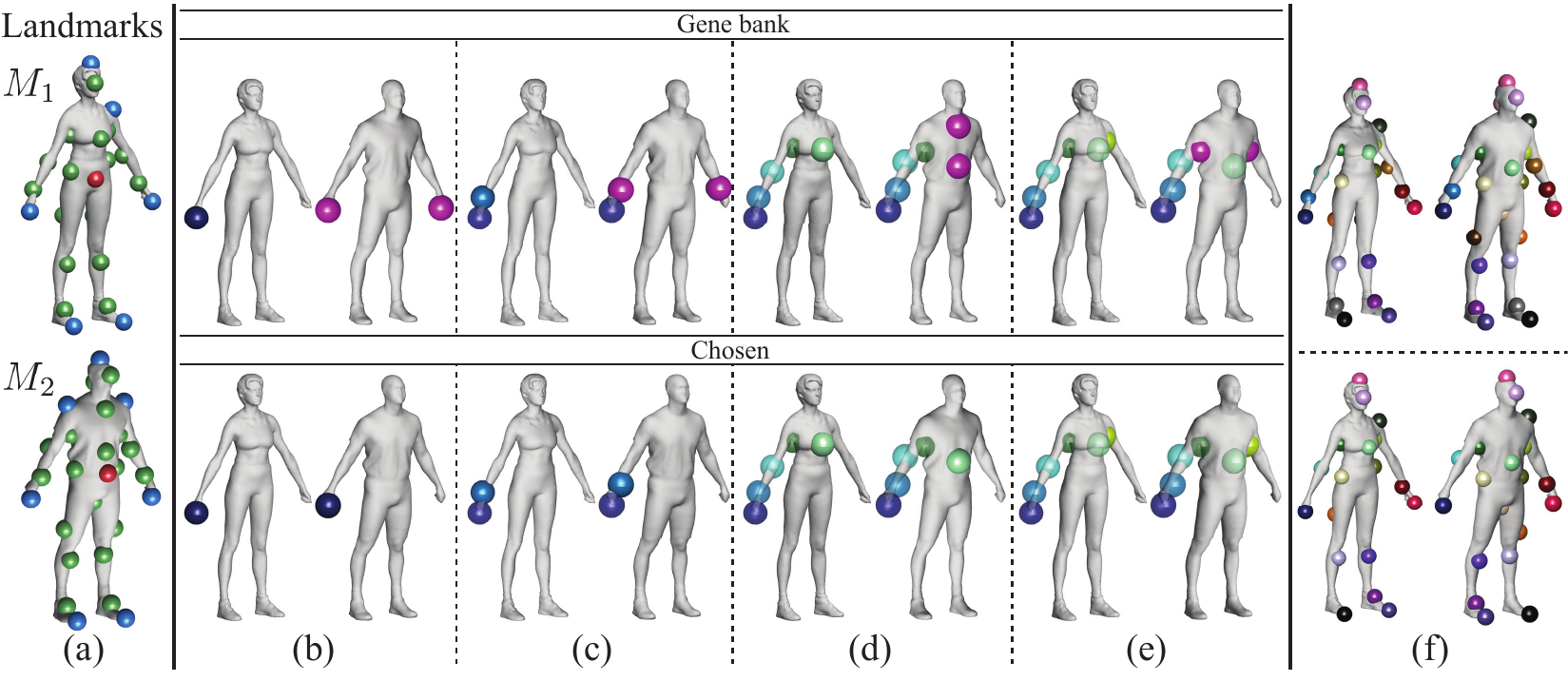}
		\caption{Random chromosome construction. (a) Input landmarks, colored by origin. (b)-(e) iterative addition of genes to the chromosome. The top row shows the landmark on the source, and the gene bank potential matches on the target (in magenta); the bottom row shows the chosen pair of a gene in matching colors. (f) (top) chromosome after adding all the genes and (bottom) after adjusting the match size. See the text for details.} 
		\label{fig:initDem}
	\end{figure*}

	Using these definitions, we can now address the construction of a new  chromosome $\chrom$, as seen in Algorithm~\ref{alg:init} and demonstrated in Figure~\ref{fig:initDem}.
	
	First, we randomly select a \emph{prominent landmark} $l_1$, and add a random 
	gene from its gene bank $\geneBank(l_1)$ to $\chrom$. 
	We maintain two subsets $\landSetAss_1, \landSetUnProc_1 \subseteq \landSet_1$, 
	that denote the landmarks that have non-empty genes in $\chrom$, and the 
	unprocessed landmarks, respectively. Hence, initially, $\landSetAss_1 = 
	\{l_1\}$, and $\landSetUnProc_1 = \landSet_1 \setminus \{l_1\}$. 
	
	Then, we repeatedly find a \emph{closest matching pair} $[\landAss_1, 
	\landUnProc_1]$, and try to add an \emph{adjacency preserving gene} for 
	$\landUnProc_1$ which keeps the chromosome \emph{valid}. First, we look for an AP 
	gene in the gene bank $\geneBank(\landUnProc_1)$. If none is found, we try to 
	construct an AP gene $(\landUnProc_1, l_2)$, where $l_2 \tin \landSet_2$, and is 
	of the same origin as $\landUnProc_1$.
	
	If no AP gene can be constructed which maintains the validity of $\chrom$, an 
	empty gene for $\landUnProc_1$ is added to $\chrom$, and we look for the next 
	closest matching pair.
	If $\adjSet(\landSetAss_1) \cap \landSetUnProc_1$ is empty, namely, no more 
	adjacent landmarks remain unmatched, empty genes are added to $\chrom$ for all 
	the remaining landmarks in $\landSetUnProc_1$.

	Figure~\ref{fig:initDem} demonstrates the chromosome construction. Both meshes and their computed landmarks (color indicates landmark origin) are displayed in (a). In the first stage, a random \emph{prominent landmark} is selected, its gene bank (GB) is displayed in magenta on $\Tar$ (b, top). Then, one random feature is selected from the GB (b, bottom). In the next stage, the \emph{closest matched/unmatched pair} is selected, and its matching options from the GB are displayed in magenta (c, top). Since only the option from the first chromosome is \emph{adjacency preserving (AP)}, this option is selected (c, bottom). After adding two more landmarks in a similar fashion, the landmark on the chest has a few \emph{AP} and unselected matching options in the GB (d, top), therefore a random landmark is selected (d, bottom). The next \emph{closest unmatched landmark} has two matching options (e, top), both are \emph{AP}, yet the one under the right armpit was previously selected. Therefore the landmark under the left armpit is chosen (e, bottom). After going through all the landmarks, the resulting chromosome is displayed (f, top). After enforcing the \emph{match size} (see Section~\ref{subsec:matchSize}), the resulting chromosome is displayed in (f, bottom).
	
	\subsubsection{Match size}
	\label{subsec:matchSize}
	Given a chromosome $\chrom$, the number of non-empty genes $|\Pi(\chrom)|$ determines 	how many landmarks are used for computing the dense map. On the one hand, if $|\Pi(\chrom)|$ is too small, it is unlikely that the computed map will be useful. On the other hand, we want to have a variety of possible maps, and thus we require the number of non-empty matches to vary in size. 
	
	Hence, before a chromosome is constructed, we randomly select a match 
	size $m_{\min} \leq \matchSize \leq m_{\max}$, where $m_{\max}$ is bounded by the landmarks sets' size $m_1, m_2$, and $m_{\min}$ is a constant fraction of it.

	To enforce the match size to be $\matchSize$, we discard the constructed chromosome if it does not have enough non-empty genes. If it has too many non-empty genes, we remove genes randomly until we reach 
	the required size. We only remove genes that originate from centers landmarks, as they are often less salient than the other two landmark classes.

	\subsubsection{Population construction}
	We construct chromosomes as described previously until completing the initial population or a maximal iteration is reached. During construction, repeated chromosomes are discarded.

	\begin{algorithm}[b]
		\SetKwInOut{Input}{input}\SetKwInOut{Output}{output}
		\SetKwProg{Fn}{Function}{ is}{end}
		
		\Input{Landmark adjacencies $\adjSet$, Landmark origins $\origSet$}
		\Output{A chromosome $\chrom$}
		Pick a random prominent landmark $l_1 \in \landSet_1$ \tcp*[r]{seed 
			landmark}
		Add a random gene from $\geneBank(l_1)$ to $\chrom$ \tcp*[r]{first gene 
			from gene bank}
		$\landSetAss_1 = \{l_1\}, \landSetUnProc_1 = \landSet_1 \setminus \{l_1\}$ 
		\tcp*[r]{initialize sets}
		\While(\tcp*[f]{Adjacent unmatched landmarks 
			exist}){$\adjSet(\landSetAss_1) \cap \landSetUnProc_1 \neq \emptyset$} { 
			Find closest matched/unmatched pair $[\landAss_1, \landUnProc_1]$ \\
			$g^{+}$ = $(\landAss_1, \chrom\chromTarOf{\landAss_1})$ \tcp*[r]{adjacent 
				gene}
			$g$ = pickGene($c, g^{+}, \landUnProc_1, \{\geneBank(\landUnProc_1), 
			\origSet(\landUnProc_1)\}$) \tcp*[r]{gene to add}
			add $g$ to $c$ \\
			remove $\landUnProc_1$ from $\landSetUnProc_1$ \tcp*[r]{landmark was 
				processed}
			\If{$g$ is not empty}{
				add $\landUnProc_1$ to $\landSetAss_1$
			}
		}
		add empty genes for all $\landUnProc_1 \in \landSetUnProc_1$  
		\tcp*[r]{remaining unmatched landmarks}
		\algspace
		\Fn{pickGene($c, g^{+}, \landUnProc_1, \landSetUnProc_2$)}{
			\ForEach{$\landUnProc_2 \in \landSetUnProc_2$}{	
				$g$ = $(\landUnProc_1, \landUnProc_2)$ \\
				\If{$g$ is AP to $g^{+}$ and $c \cup g$ is valid}{
					return $g$ \\
				}
			}
			return empty gene \\	
		}
		\caption{Create a random chromosome.\vspace{-.2in}}\label{alg:init}
	\end{algorithm}		
	
	\subsection{Fitness}
	\label{sec:fitness}
	
	\begin{figure}[t!]
		\centering
		\includegraphics[width=\linewidth]{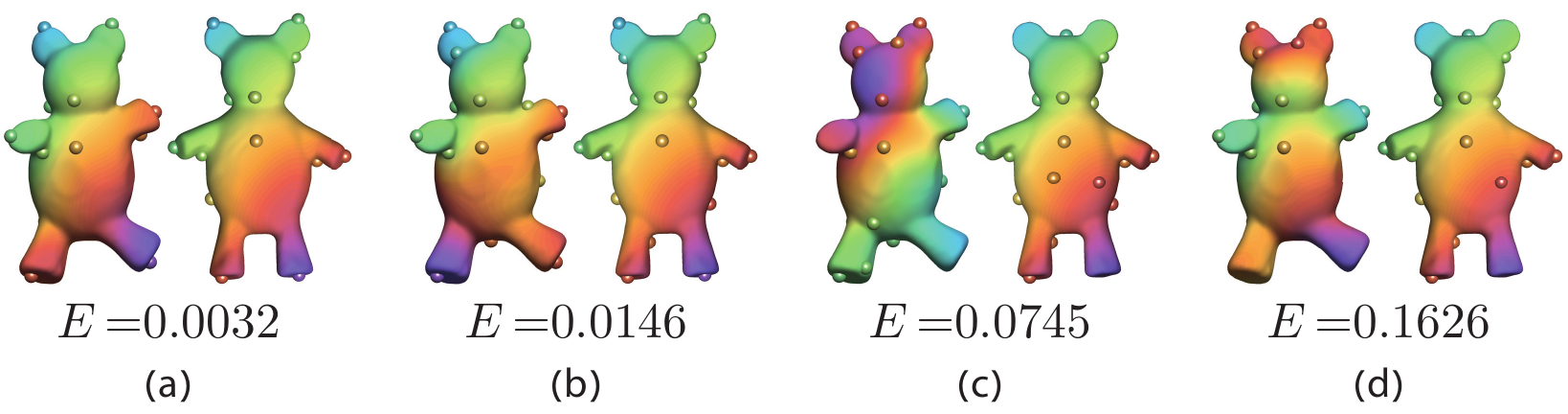}
		\caption{\CH{The reversible elastic energy of a few chromosomes. (a) The correct match, (b) the legs are switched, (c) the head and legs are switched (d) an incorrect match. Note that the enetgy gradually increases as more landmarks are mapped incorrectly.}} 
		\label{fig:EngBhv}
	\end{figure}

The fitness of a chromosome $\chrom$ is evaluated by extracting a functional map from its match $\Pi(\chrom)$, and evaluating its fitness energy.	
The \emph{fittest} chormosome is the one with the \emph{lowest} fitness energy.
	
	\paragraph*{Functional map optimization.}
	The match $\Pi(\chrom)$ defines a permutation that maps the subset $\landSetAss_1$ to $\landSetAss_2$ and \emph{vice versa}. Thus we use it to compute functional maps in \emph{both} directions, by optimizing 
	\begin{equation}
	\hat{\fmap}_{ij}(\Pi(\chrom)) = \underset{\fmap_{ij}}{\text{argmin}} \,\, 
	\Efmap(\fmap_{ij}, \Pi(\chrom)),
	\label{eq:hatfmap}
	\end{equation}
	where $(i,j) \tin \{(1,2), (2,1)\}$ and $\Efmap$ is given in Equation~\eqref{eq:fmap}. This 
	computation is very efficient, as these are two unconstrained linear least 
	squares problems.

	\paragraph*{Functional map refinement.}
	The optimized functional maps $\hat{\fmap}_{ij}$ are efficiently refined 
	by converting them to pointwise maps and back, such that they better represent 
	valid pointwise maps. We solve:
	\begin{equation}
	{\pwiseMat}_{ij}(\hat{\fmap}_{ij}) = \underset{\pwiseMat}{\text{argmin}} \,\, 
	\|\pwiseMat \vertexCoords_j - \pwiseMat(\hat{\fmap}_{ij}) \vertexCoords_j\|^2_F,
	\label{eq:pmap}
	\end{equation}
	where $P$ is a binary row-stochastic matrix, and $\pwiseMat(\fmap_{ij})$ is 
	given in Equation~\eqref{eq:fmap_to_pwise}. This problem can be solved 
	efficiently using a $kd$-tree, by finding nearest neighbors in $\xR^3$.
	Finally, the optimal functional maps in both directions are given by:
	$\fmap_{ij}^{\,\text{opt}} = \fmap(\pwiseMat_{ij})$, 
	where $\fmap(\pwiseMat_{ij})$ is given in Equation~\eqref{eq:pwise_to_fmap}.
	
	\begin{figure*}[t!]
		\centering
		\includegraphics[width=\linewidth]{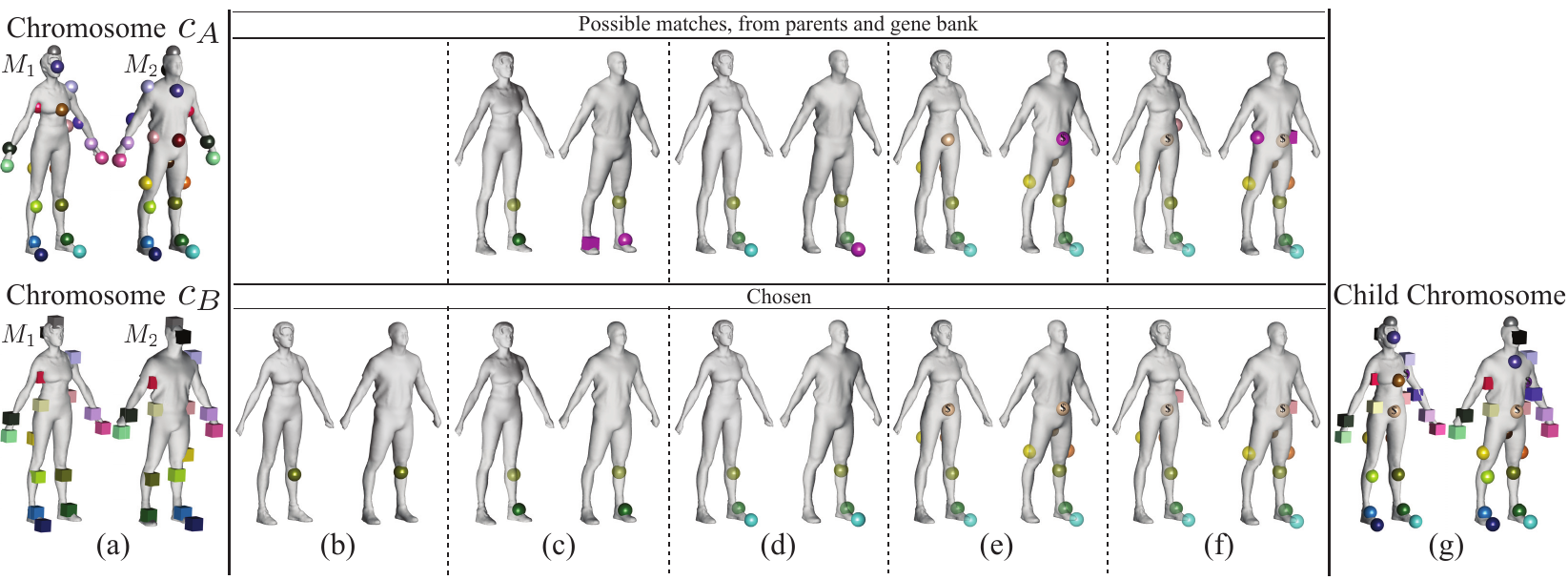}
		\caption{The crossover of two chromosomes $\chrom_A, \chrom_B$ (a) yields a new child chromosome (g). (b-f) the top row shows in each step the potential genes on the target (magenta) with a landmark on the source, and the bottom row shows the chosen genes. (b) initial seed from $\chrom_A$, (c) closest unmatched landmark, gene chosen by adjacency, (d) only one potential gene, (e) no potential genes from parents, gene picked from gene bank (marked with $\$$). (f) gene chosen by adjacency. Note that $\chrom_A$ switches the hands, $\chrom_B$ switches the legs, and the child chromosome matches both correctly.
		} 
		\label{fig:crossover}
	\end{figure*}

	\paragraph*{Functional map fitness.}
	Finally, we evaluate the fitness of the maps with the \emph{reversible elastic 
		energy}:
	\begin{equation}
	\Efit (\fmap_{12}^{\,\text{opt}}, \fmap_{21}^{\,\text{opt}} ) = \gamma 
	\sum_{ij} \Eelastic (\fmap_{ij}^{\,\text{opt}})  + (1-\gamma ) \, \Erevers 
	(\fmap_{12}^{\,\text{opt}},\ \fmap_{21}^{\,\text{opt}} ) ,
	\end{equation}
	where $\Eelastic,\ \Erevers$ are defined in Equations~\eqref{eq:elastic} and~\eqref{eq:rev}, respectively. 
	While this objective is highly non-linear and non-convex, the fitness is never 
	optimized directly, but only evaluated during the genetic algorithm, which is 
	well suited for such complex objective functions.
	\CH{Figure~\ref{fig:EngBhv} demonstrates the behavior of the fitness function for a few chromosomes. The sparse correspondences and
	the resulting functional maps of four chromosomes are displayed. The \emph{reversible elastic energy} of the correct match is the lowest (a), while a small change such as switching only the legs (b) results in a higher energy. When all the landmarks are mapped incorrectly yet in a consistent way as in (c), where the head and legs are reversed, the energy is even higher, and the highest energy is obtained when the landmarks are mapped incorrectly and inconsistently (d).}

	\subsection{Selection}
	\label{sec:selection}
	
	In this process, individuals from the population are selected in order to pass 
	their genes to the next generation. At each stage half of the population is 
	selected to mate and create offspring\CH{s}. In order to select the individuals for 
	mating we use a \emph{fitness proportionate selection}~\cite{back1996evolutionary}. In our case the probability to select an 
	individual for mating is proportional to $1$ over its fitness, so that fitter 
	individuals have a better chance of being selected. 
	
	\begin{algorithm}[b]

		\SetKwInOut{Input}{input}\SetKwInOut{Output}{output}
		\SetKwProg{Fn}{Function}{ is}{end}
		
		\Input{Landmark adjacencies $\adjSet$, chromosomes $\chrom_A, \chrom_B$}
		\Output{Chromosomes $\chromChild_A, \chromChild_B$}
		$\landSetAss_{1A} = \landSetAss_1(\Pi(\chrom_A)), \, \landSetAss_{1B} = 
		\landSetAss_1(\Pi(\chrom_B))$ \\
		$p(\chromChild_A) = \chrom_A, \, p(\chromChild_B) = \chrom_B$ 
		\tcp*[r]{set parents}
		Pick a random landmark $l_1 \in \landSetAss_{1A} \cap \landSetAss_{1B}$ 
		\tcp*[r]{seed landmark}
		\ForEach{$\chromChild \in \{\chromChild_A, \chromChild_B\}$}{
			Copy the $l_1$ gene from $p(\chromChild)$ to $\chromChild$ 
			\tcp*[r]{first gene}
			$\landSetAss_1 = \{l_1\}, \landSetUnProc_1 = \landSet_1 \setminus 
			\{l_1\}$ \tcp*[r]{initialize sets}
			\While(\tcp*[f]{unmatched landmarks exist}){$\landSetUnProc_1 \neq 
				\emptyset$} { 
				\eIf(\tcp*[f]{Adjacent landmarks exist}){$\adjSet(\landSetAss_1) 
					\cap \landSetUnProc_1 \neq \emptyset$} { 
					Find closest matched/unmatched pair $[\landAss_1, \landUnProc_1]$ \\
					$g^{+}$ = $(\landAss_1, \chromChild\chromTarOf{\landAss_1})$ 
					\tcp*[r]{adjacent gene}
					$\mathcal{P}$ = $\{(\landUnProc_1, 
					\chrom_A\chromTarOf{\landUnProc_1}), (\landUnProc_1, 
					\chrom_B\chromTarOf{\landUnProc_1})\}$ \tcp*[r]{parent genes}
					$g$ = pickGene($\chromChild, g^{+}, \landUnProc_1, \{\mathcal{P}, 
					\geneBank(\landUnProc_1)\}$) \tcp*[r]{gene to add}
				}{
					$g$ = random gene from $\Pi(p(\chromChild))$ such that 
					$\chromChild \cup g$ is valid \\
					\If(\tcp*[f]{no more parent genes}){$g = \emptyset$}{
						break \\
					}
					$\landUnProc_1$ = source landmark of $g$ \\
				}
				add $g$ to $\chromChild$ \\
				remove $\landUnProc_1$ from $\landSetUnProc_1$ \tcp*[r]{landmark was 
					processed}
				\If{$g$ is not empty}{
					add $\landUnProc_1$ to $\landSetAss_1$
				}	
			}
		}
		add empty genes for all $\landUnProc_1 \in \landSetUnProc_1$  
		\tcp*[r]{remaining unmatched landmarks}
		\caption{Crossover.\vspace{-.2in}}
		\label{alg:crossover}
	\end{algorithm}

	\subsection{Crossover}
	\label{sec:crossover}

	The chromosomes selected for the next generation undergo a \emph{crossover} operation with probability \pcross.
	The crossover operator merges two input chromosomes $\chrom_A, 
	\chrom_B$ into two new chromosomes $\chromChild_{A}, \chromChild_{B}$. 
	To combine the input chromosomes in a geometrically consistent way, we again 
	use adjacency preserving (AP) genes, as defined in Section~\ref{sec:genes}. The algorithm is similar to the 
	initial chromosome creation, however, rather than matching landmarks based on the gene bank alone, matching is mainly based on the \emph{parent} chromosomes. 
	
	The crossover algorithm is described in Algorithm~\ref{alg:crossover}.
	First, we randomly pick a non empty gene from the parent chromosomes. The correspondence of the selected gene as assigned by $\chrom_A$ is the seed of $\chromChild_{A}$, and correspondence of the same gene as assigned by $\chrom_B$ is the seed of $\chromChild_{B}$.
	Then, each chromosome is constructed by iteratively adding valid AP genes from the parents. If all the AP genes invalidate the child chromosome, we consider the gene bank options instead, and if these are not valid as well, we pick a random gene that does not invalidate the child chromosome.

	Figure~\ref{fig:crossover} demonstrates the creation of $\chromChild_{A}$ from the parent chromosomes shown in (a). We show corresponding landmarks in the same color, and potential assignments (from the parents or from the gene bank) in magenta. The shape of the landmark (sphere or cube) indicates the parent chromosome. 
	
	First, a random non empty gene is copied from $\chrom_A$ (b). Then, the \emph{closest unmatched landmark} is selected (we show the potential genes of both parents) (c, top). Since only the gene of $\chrom_A$ is AP, this gene is selected (c, bottom). In (d), the next closest unmatched landmark has only one potential gene, since only $\chrom_A$ has a non empty gene with this landmark. Since this option is AP and does not invalidate the chromosome, it is selected. After matching a few more landmarks, a landmark on the stomach is chosen (e, top). This landmark has no potential genes from the parents, therefore, the potential genes are taken from the gene bank (marked with $\$$). In this case there are two possible matching options, one is visible on the stomach, and another is on the back of $\Tar$ and is not visible in the figure. Both options are AP and do not invalidate the chromosome, so one is chosen randomly (e, bottom). The next closest unmatched landmark has two potential genes (f, top). Both are AP and valid, therefore, one is selected randomly (f, bottom). 
	
	After repeating this process for all source landmarks, the resulting chromosome is shown (g).  $\chrom_A$ maps the legs correctly, yet the arms are switched, whereas $\chrom_B$ switches the legs but correctly matches the arms.  
	The child chromosome is, however, better then both its parents, since both the legs and the hands are mapped correctly.

	\begin{figure*}[t!]
		\centering
		\includegraphics[width=0.83\linewidth]{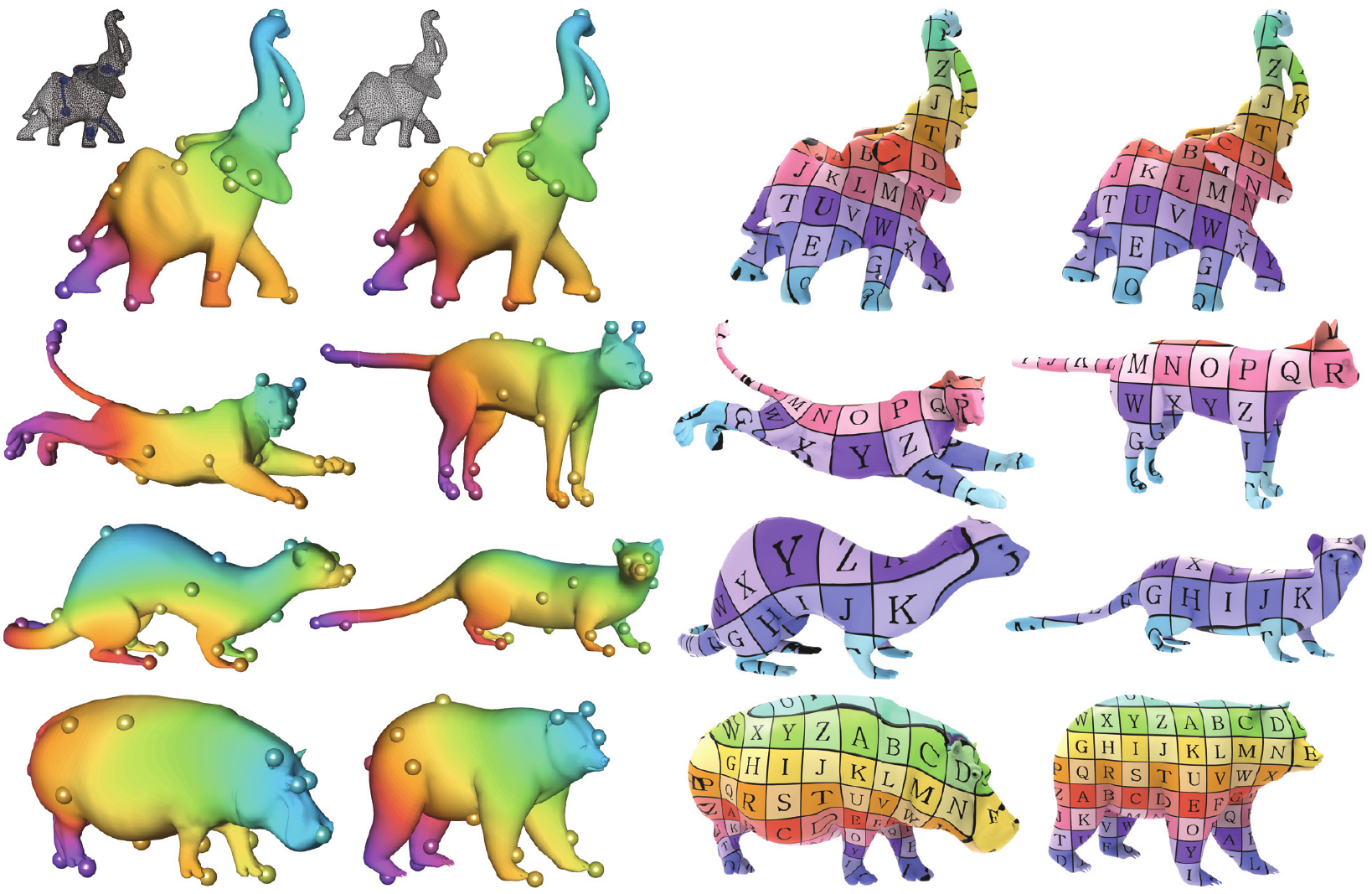}
		\caption{\newT{Some qualitative results of our algorithm on meshes of different genus (top), and non isometric shapes (center, bottom).} } 
		\label{fig:newWINres}
	\end{figure*}

	\subsection{Mutation}
	\label{sec:mutation}
	
    A new chromosome $\chrom$ can undergo three types of mutations, with some mutation-dependent probability. We define the following mutation operators~\cite{brie2005genetic}.

	\subsubsection{Growth (probability \pmutG).} Go over all the empty 
	genes in random order and try to replace one, $(l_1, 0)$, by assigning it a 
	corresponding landmark using the following two options.
	\paragraph*{FMap.} Compute $P_{12}(\hat{C}_{12}(\Pi(\chrom)))$, using 
	equations~\eqref{eq:hatfmap} and~\eqref{eq:pmap} and set $l_2$ to the closet 
	landmark on $M_2$ to $P_{12}(l_1)$. Add $g = (l_1, l_2)$ to $\chrom$ if $\chrom 
	\cup g$ is valid.
	\paragraph*{Gene Bank.} If the previous attempt failed, set $g$ to a random 
	gene from the gene bank $\geneBank(l_1)$, and add it to $\chrom$ if $\chrom 
	\cup g$ is valid.

	\subsubsection{Shrinkage (probability \pmutS).} Randomly select 
	$n_{\text{sh}}$ \emph{centers} landmarks from $\landSetAss_1(\Pi(c))$.
	Replace each one of the corresponding genes with an empty gene, such that 
	$n_{\text{sh}}$ new chromosomes are obtained. Each of these has a single new 
	empty gene. The result is the fittest chromosome among them (including the 
	original).
	
	\subsubsection{FMap guidance (probability \pmutFMG).} 
	Using equations~\eqref{eq:hatfmap} and~\eqref{eq:pmap}, compute $P_{12}(\hat{C}_{12}(\Pi(\chrom)))$. For each landmark 
	$l_1\tin\landSetAss_1(\Pi(c))$ set $l_2$ to the closet landmark on $M_2$ to 
	$P_{12}(l_1)$. If injectivity is violated, i.e. $\chrom \chromTarOf{i} = \chrom 
	\chromTarOf{j}$, if $i$ is a center and $j$ is in another category, set $\chrom 
	\chromTarOf{i} = 0$ and $\chrom \chromTarOf{j}$ as computed by the functional 
	map guidance. 
	Otherwise, randomly keep one of them and set the other one to 0. We prioritize 
	maxima and minima as they are often more salient.

	\subsection{Stopping criterion}
	\label{sec:convergence}
	We stop the iterations when the fittest chromosome remains unchanged for a set amount of iterations, meaning the population has converged, or when a maximal iteration number is reached.

	\subsection{Limitations}

	\begin{wrapfigure}[6]{r}{0.3\columnwidth}
		\vspace{-.025\textheight}
		\centering
		\hspace*{-15pt}
		\includegraphics[width=1.1\linewidth]{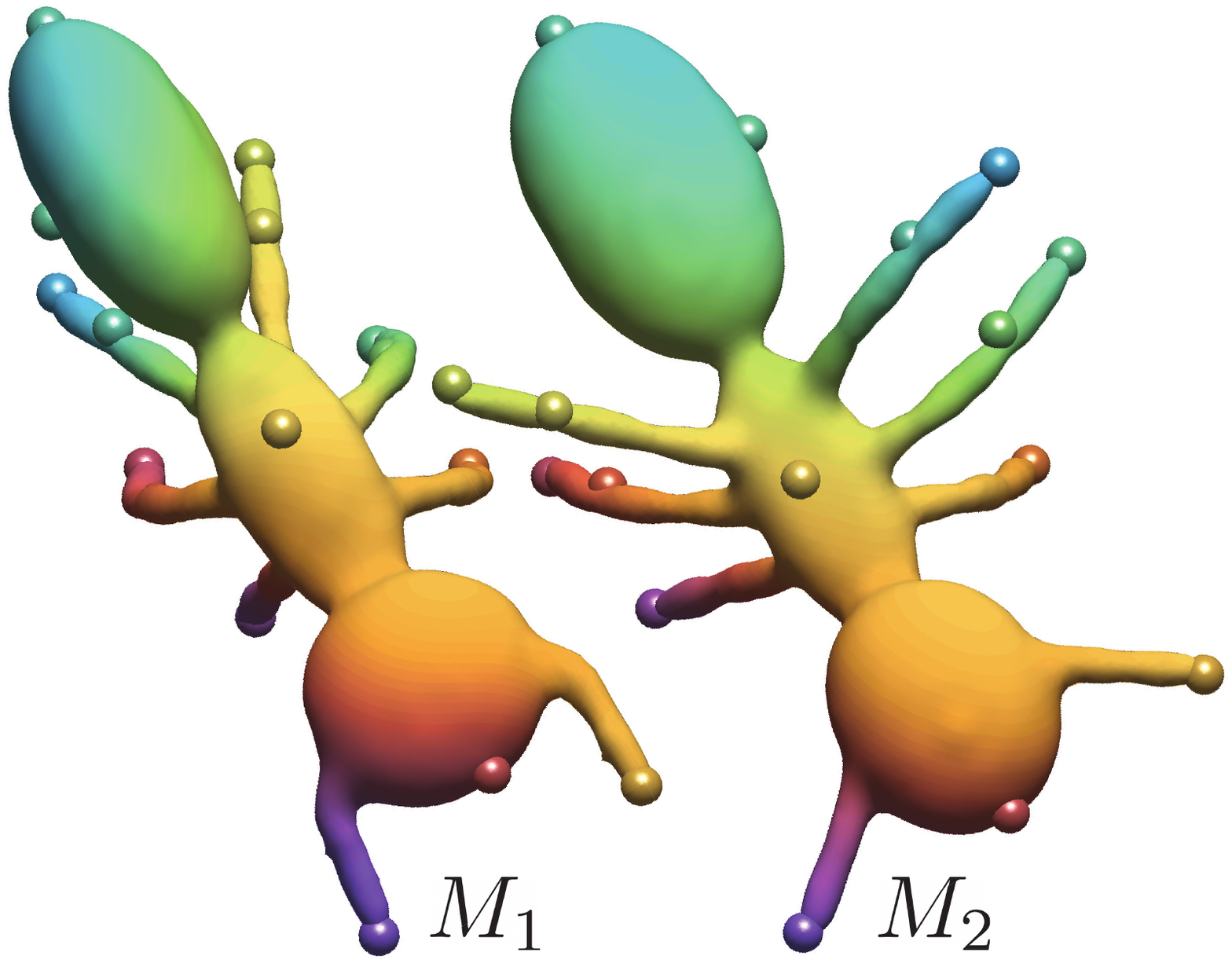}
	\end{wrapfigure}

	We evaluate the quality of a sparse correspondence using the elastic energy of the induced functional map, which is represented in a reduced basis. In some cases, the reduced basis does not contain enough functions to represent correctly the geometry of thin parts of the shape, and as a result both correct and incorrect sparse correspondences can lead to a low value for the elastic energy. Such an example is shown in the inset Figure. Both the body and the legs of the ants are very thin, and the resulting best match switches between the two back legs.

	\subsection{Timing}
	The most computationally expensive part of the algorithm is computing the fitness of each chromosome. While computing the functional map and evaluating the elastic energy of one chromosome is fast, this computation is done for all the offspring in each iteration, and is therefore expensive.
	Our method is implemented in MATLAB. On a desktop machine with an Intel Core i7 processor, for meshes with $5$K vertices, computation of a functional map for a single chromosome takes $0.035$ seconds and the computation of the elastic energy takes $0.002$ seconds. The average amount of iterations it took for the algorithm to converge is $225$ and the average total computation time is $10.5$ minutes.

	\section{Results}
	\label{sec:results}
	
	\begin{figure*}[t!]
		\centering
		\includegraphics[width=0.95\linewidth]{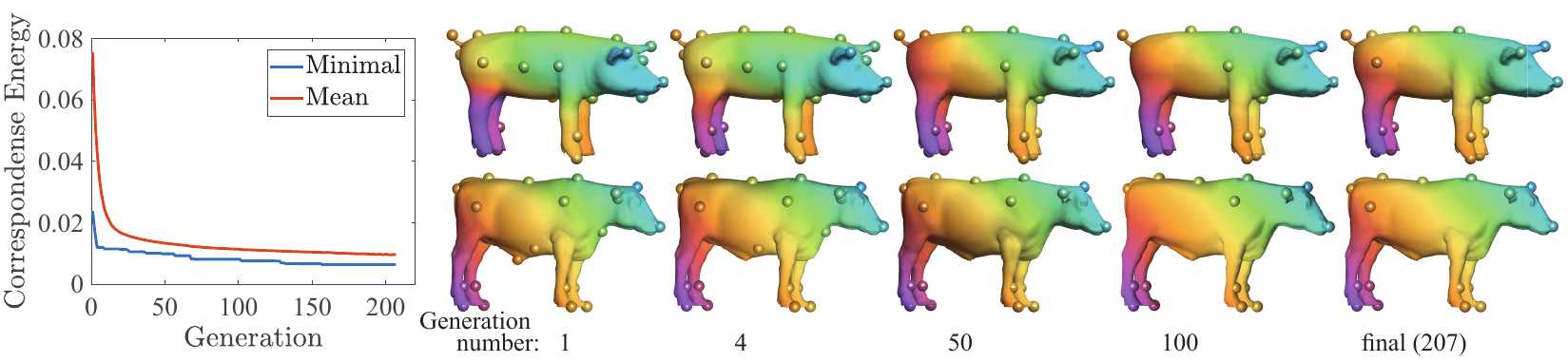}
		\caption{\newT{Population evolution.} (left) The energy of the fittest offspring and the mean energy of the population decrease until convergence indicating improvement throughout the generations. (right) The fittest offspring at different generations, improving until reaching the \newT{correct correspondence}.} 
		\label{fig:convergence}
	\end{figure*}

	\subsection{Datasets and Evaluation}
	\label{sec:eval_proto}
	Our method computes a sparse correspondence and a functional map that can be 
	used as input to existing semi-automatic methods such as~\cite{RHM}, that we 
	use in this paper. 
	
	\paragraph{Datasets.}
	We demonstrate the results of our method on two datasets with different 
	properties.
	The SHREC'07 dataset~\cite{shrec07} contains a variety of non-isometric shapes 
	as well as ground truth sparse correspondence between manually selected 
	landmarks. This dataset is suitable to demonstrate the advantages of our 
	method, since we address the highly non-isometric case.
	The recent dataset SHREC'19~\cite{SHREC19} contains shapes of the same semantic 
	class but different topologies, that we use to demonstrate our results in such 
	challenging cases. 
	
	\paragraph{Quantitative evaluation.}
	Quantitatively evaluating sparse correspondence on these datasets is challenging, 
	since the given sparse ground truth does not necessarily coincide with the 
	computed landmarks. We therefore quantitatively evaluate the results after post 
	processing, where we use the same post processing for all methods (if a method 
	produces sparse correspondence we compute functional maps using these landmarks 
	and run RHM~\cite{RHM} to extract a pointwise map, and if a method produces a 
	pointwise map we apply RHM directly). Since we initialize RHM with a functional 
	map or a dense map, we use the Euclidean rather than the geodesic embedding 
	that was used in their paper, that is only needed when the initialization is 
	very coarse.
	We use the evaluation protocol suggested by Kim et al.~\shortcite{kim2011blended}, 
	where the $x$ axis is a geodesic distance between a ground truth correspondence 
	and a computed correspondence, and the $y$ axis is the percentage of 
	correspondences with less than $x$ error.
	We also allow symmetries, as suggested by Kim et al.~\shortcite{kim2011blended}, by 
	computing the error w.r.t. both the ground truth correspondence and the 
	symmetric map (that is given), and using the map with the lower error as ground 
	truth for comparison.

	\paragraph{Qualitative evaluation.}
	We qualitatively evaluate our results using color and texture transfer.
	We visualize a sparse correspondence by showing corresponding landmarks in the 
	same color.
	To visualize functional maps we show a smooth function, visualized by color 
	coding on the target mesh, and transfer it to the source using the functional 
	map.
	We visualize pointwise maps by texture transfer.
	\newT{Figure~\ref{fig:newWINres} demonstrates a few examples of our results using these visualizations.}

	\subsection{Population Evolution and Convergence}
	
	Figure~\ref{fig:convergence} shows the evolution of the population in the 
	pig-cow example.
	We plot the mean energy of a chromosome as a function of the generation, 
	as well as the minimal energy (of the fittest chromosome).
	We also visualize the functional map and the landmarks of the fittest 
	chromosome throughout the algorithm, which is inaccurate at the first 
	iteration (the front legs are matched incorrectly), and gets increasingly more accurate as the population evolves.

\begin{figure}[b!]
		\vspace{-.02\textheight}
		\centering
		\includegraphics[width=\linewidth]{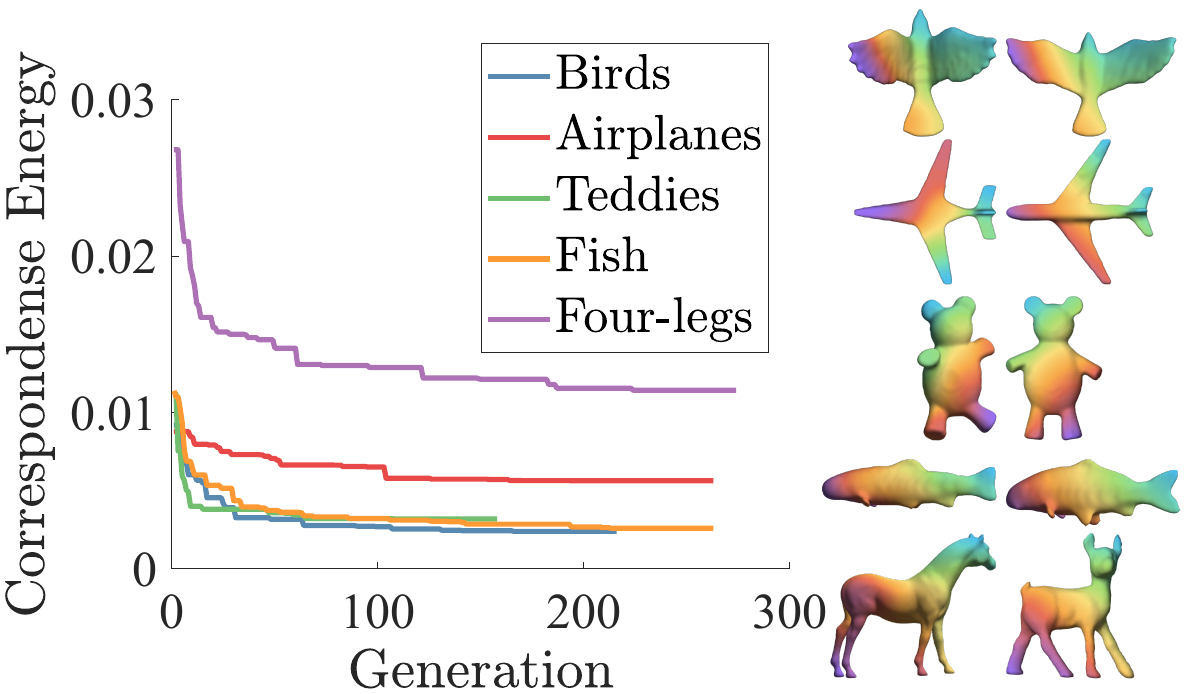}
		\caption{\newT{The correspondence energy of the fittest offspring during the iterations (left) and the final map generated from the fittest offspring in the final iteration (right) on a few pairs of shapes. Note that the energy decreases until convergence.}} 
		\label{fig:convergenceMany}
	\end{figure}

\newT{We additionally plot the energy of the best chromosome as a function of the generation number for additional shapes in Figure~\ref{fig:convergenceMany}. Similarly to the pig-cow example, the energy decreases until the algorithm converges.}

	\begin{wrapfigure}[8]{r}{0.5\columnwidth}
		\vspace{-.03\textheight}
		\centering
		\hspace{-15pt}
		\includegraphics[width=\linewidth]{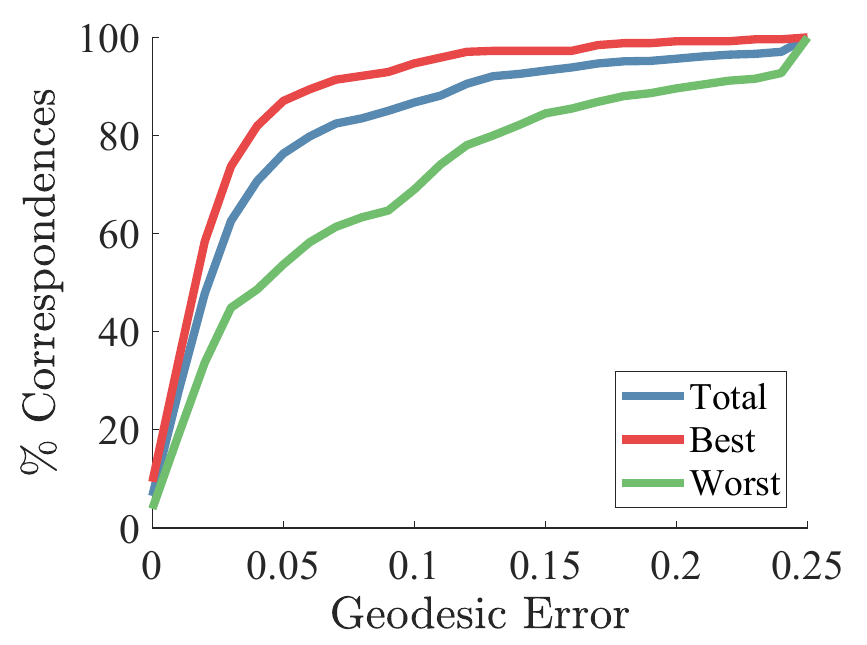}
	\end{wrapfigure}
	Since our algorithm is not deterministic, we \newT{investigate the variation of our results} by running the algorithm $100$ times on \newT{20 random pairs of shapes from the SHREC'07 dataset~\cite{shrec07}. 
	We evaluate the result using the protocol suggested by Kim et al.~\shortcite{kim2011blended}, after post processing~\cite{RHM} to extract a dense map. The results are determined by the total geodesic error with respect to the sparse ground truth and are shown in the inset Figure. The best and worst results 
 	are composed of the best and worst results of each one of the 20 pairs respectively. Note that the total results (all 100 results on all the pairs) are close to the best results, indicating that the worst results are outliers.}

	\subsection{Mapping between Shapes of Different Genus}
\begin{figure}[t!]
		\centering
		\includegraphics[width=\linewidth]{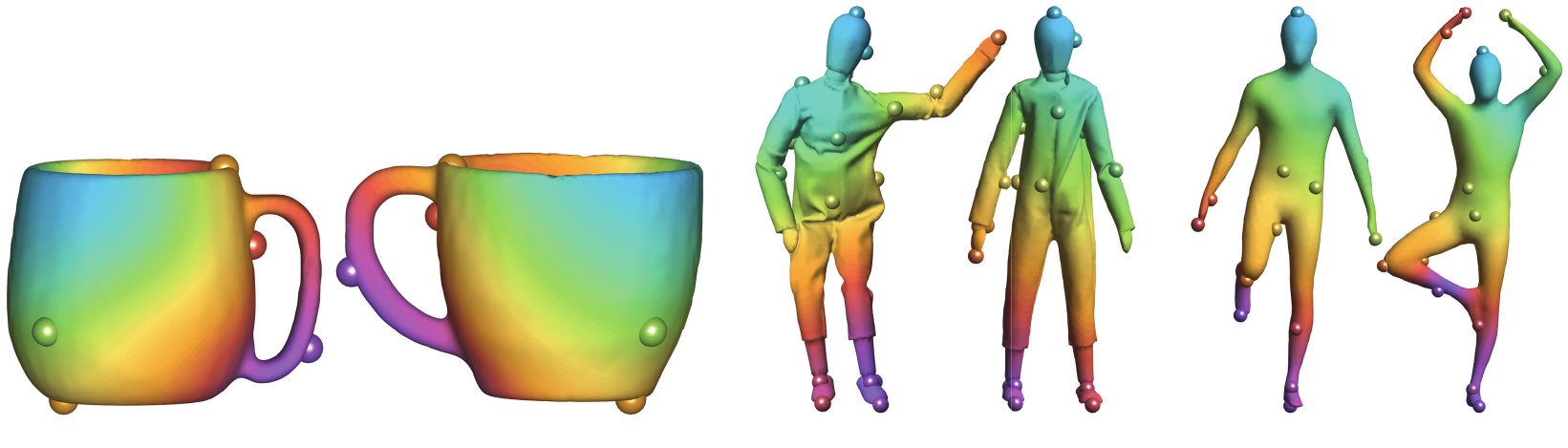}
		\caption{Our results on shapes of genus 1 (top) and shapes with 
			different genus (bottom).}
		\label{fig:high_genus}
	\end{figure}

	Our method is applicable to shapes of any genus, and can also match between shapes of different 
	topology. Figure~\ref{fig:teaser1} shows our results for two hand models from 
	SHREC'19~\cite{SHREC19}. The left model has genus $2$ 
	and the right model has genus $1$.
	Note that our approach yields a high quality dense correspondence even for such difficult cases. We additionally show in Figure~\ref{fig:high_genus} our \CH{results}
	for two cup models from 
	SHREC'07~\cite{shrec07}, both genus $1$, and two \newT{pairs of} human shapes with genus $0$ and genus $1$, respectively.
	Here as well, 
	despite the difficult topological issues, our approach \newT{computed} a meaningful map.
	\newT{Another example appears in Figure~\ref{fig:newWINres} (top). Here, we generated a genus $8$ mesh (left) by introducing tunnels through a genus $3$ mesh (right). Both meshes were then remeshed to have different triangulations. Note that even for meshes with large differences in genus, our method generates good results.}
	
	\subsection{\CH{Evaluation on increasingly non-isometric shapes.}}
\begin{figure}[t!]
		\centering
		\includegraphics[width=0.9\linewidth]{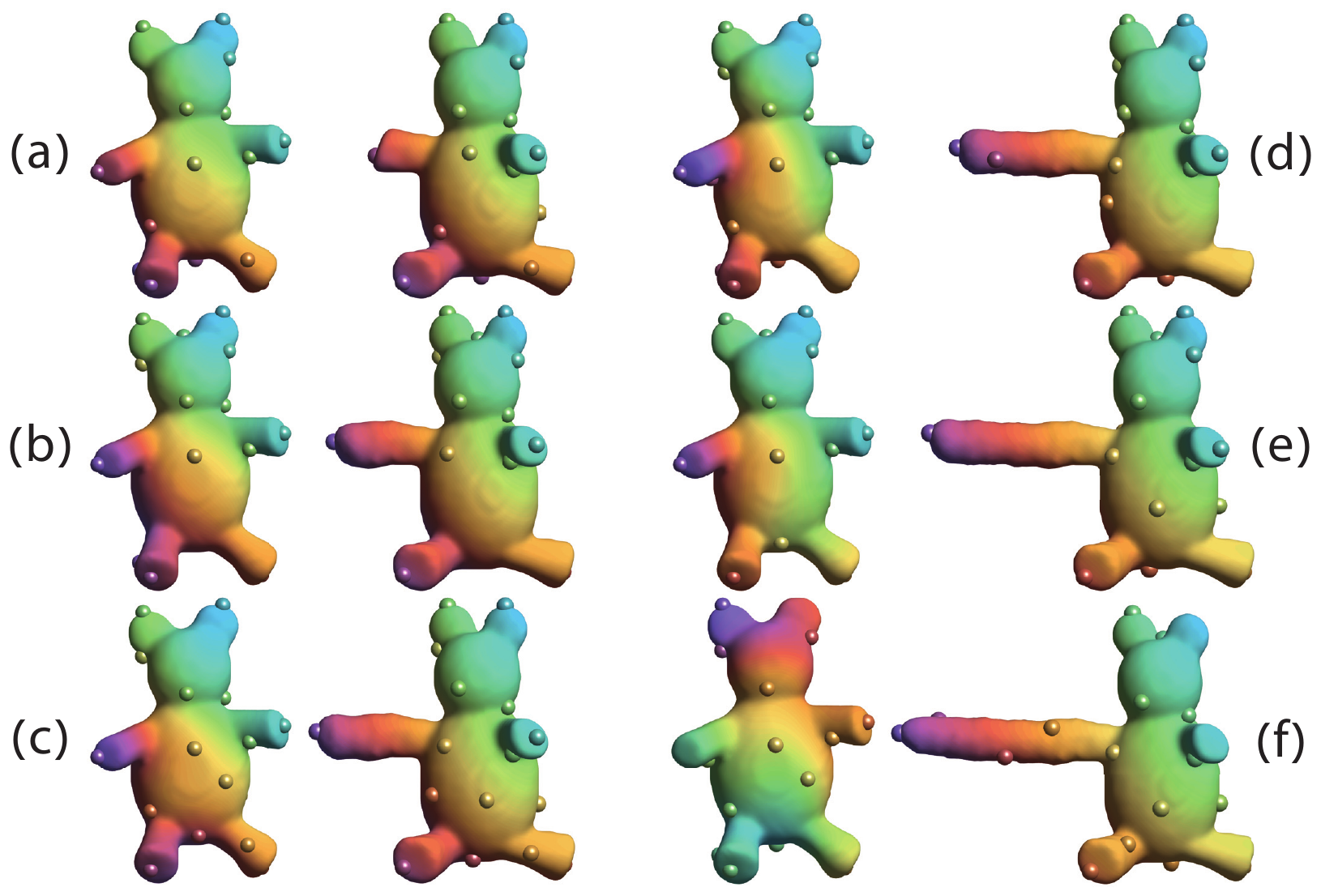}
		\caption{\CH{Varying degrees of isometry, from nearly isometric (a) to highly non isometric shapes (f). Our method finds the correct map for most pairs (a-e), yet matches incorrectly the pair with the highest distortion (f).}} 
		\label{fig:IsoTest}
	\end{figure}

	\CH{To demonstrate the efficiency of our algorithm on non-isometric shapes, we perform the following experiment. Starting from nearly isometric shapes from the SHREC'07 dataset~\cite{shrec07}, we deform one of the shapes (extend the bears' arm) gradually and obtain $5$ pairs with different degrees of isometric distortion (Figure~\ref{fig:IsoTest}). For the first $5$ pairs (a-e) we find the correct match despite the large deformation, whereas for the last pair (f), where the non-isometric deformation is very high, our algorithm does not find a correct map.}
	
	\subsection{\CH{Ablation study.}}
	\begin{wrapfigure}{r}{0.25\textwidth}
	\hspace*{-20pt}
	\centering
	\includegraphics[width=\linewidth]{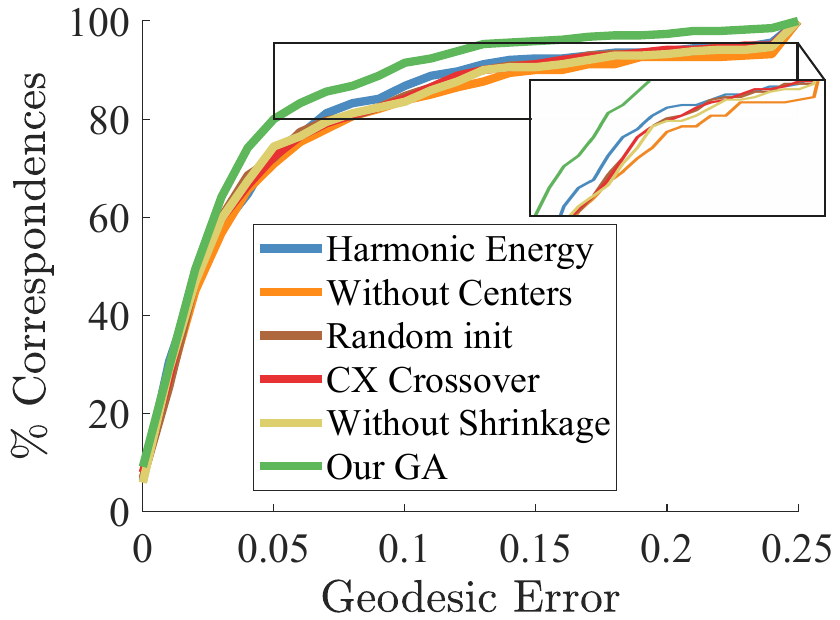}
	\label{fig:ablPartsNparams}
	\end{wrapfigure}
	
	\CH{We perform an ablation study to show the effect of different design choices of our algorithm. We run our algorithm $5$ times on $20$ random pairs from the SHREC'07 dataset~\cite{shrec07}, where each time one part of the method changes while the rest remains identical. In our first experiment, we use only the maxima and minima as landmarks, removing the centers. Next, we use a random population initialization~\cite{paul2013performance}, change the fitness to the reversible \emph{harmonic} energy~\cite{RHM} of the functional map extracted from a chromosome, use the cycle crossover (CX) from the traveling salesman problem~\cite{oliver1987study}, and finally remove the shrinkage mutation. We evaluate the results as in the previous sections and they are shown in the inset Figure. Note that changing parts of the algorithm diminishes the performance.

}

	\subsection{Quantitative and Qualitative Comparisons}
	\label{sec:QuanQual}
	
	\begin{figure}[t!]
		\centering
		\includegraphics[width=0.9\linewidth]{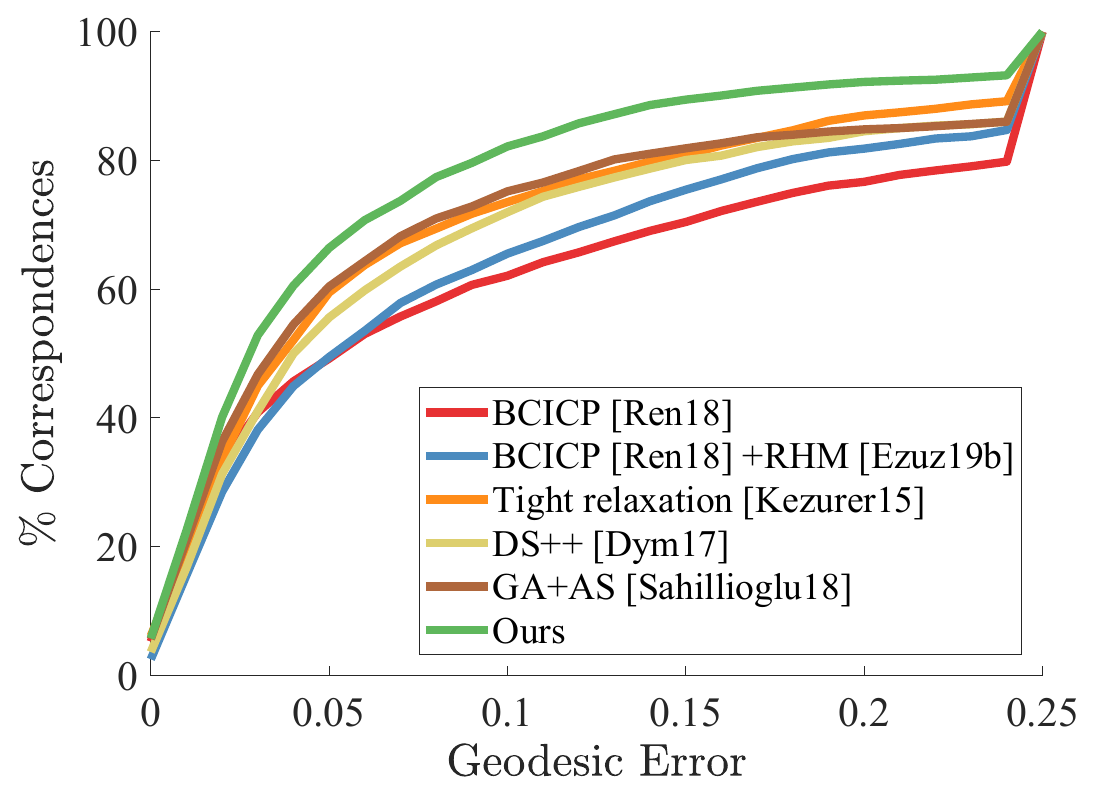}
		\caption{Quantitative comparison between our results and fully 
			automatic state of the art 
			methods~\cite{kezurer2015tight,DSppDym,SahilliogluGenetic,Ren2018Continuous}.
			We apply the same post processing~\cite{RHM} on all the methods to 
			extract a dense map, and use the evaluation protocol suggested by Kim et 
			al.~\shortcite{kim2011blended}, that measures geodesic distance to the ground truth. Note that we outperform all the other methods.} 
		\label{fig:shrec07_graphs}
	
	\end{figure}

	\begin{figure*}[t!]
		\centering
		\includegraphics[width=\linewidth]{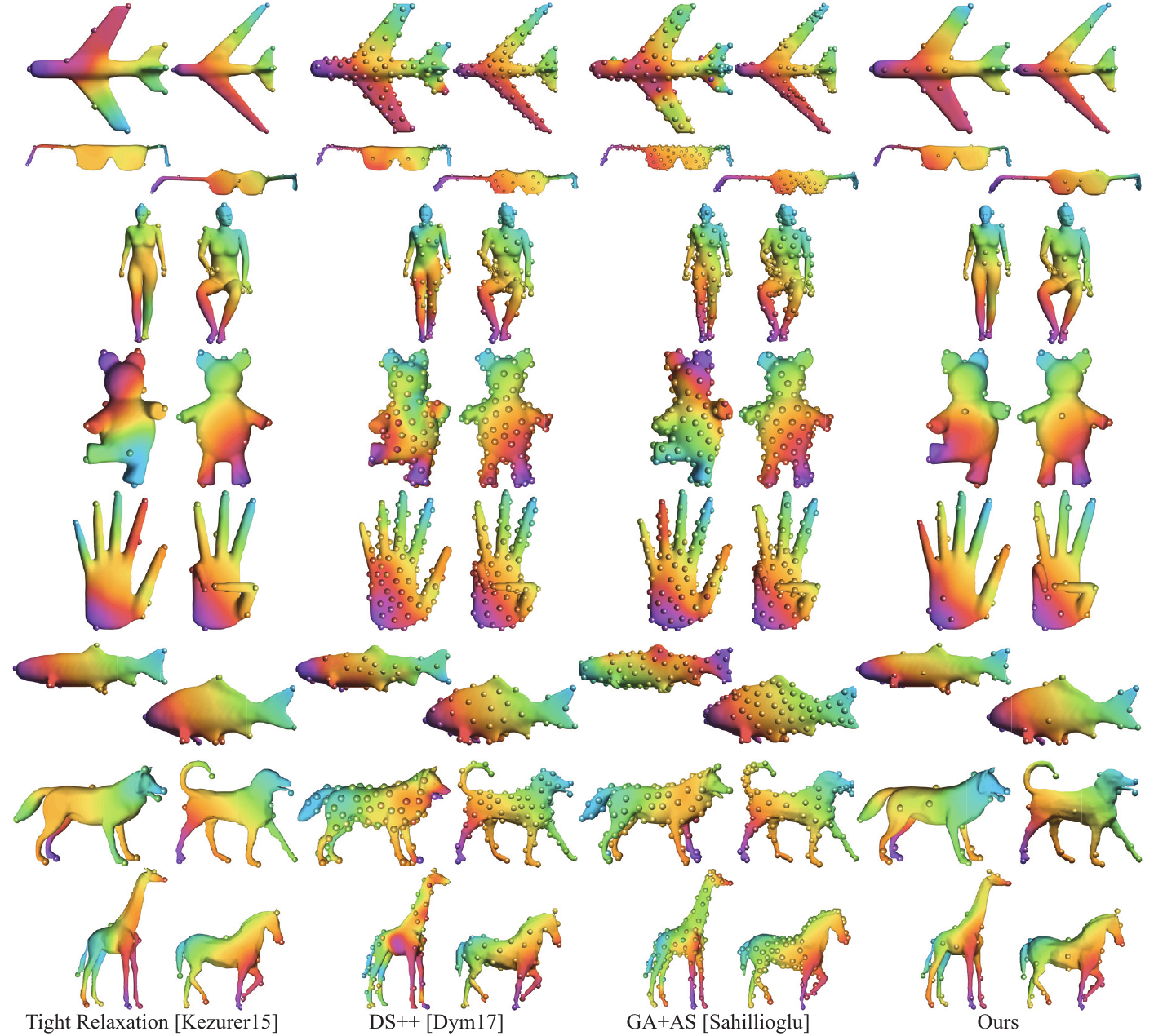}
		\caption{Qualitative comparison between the results of our genetic 
			algorithm, i.e. a sparse correspondence and a functional map, and the 
			results of automatic state of the art 
			methods~\cite{kezurer2015tight,DSppDym,SahilliogluGenetic} for sparse 
			correspondence. The functional map is visualized by color transfer (see 
			text for details), and the sparse correspondence is visualized by landmarks 
			with corresponding colors.} 
		\label{fig:shrec07_color_comp}
	\end{figure*}

	\begin{figure*}[t!]
		\centering
		\includegraphics[width=\linewidth]{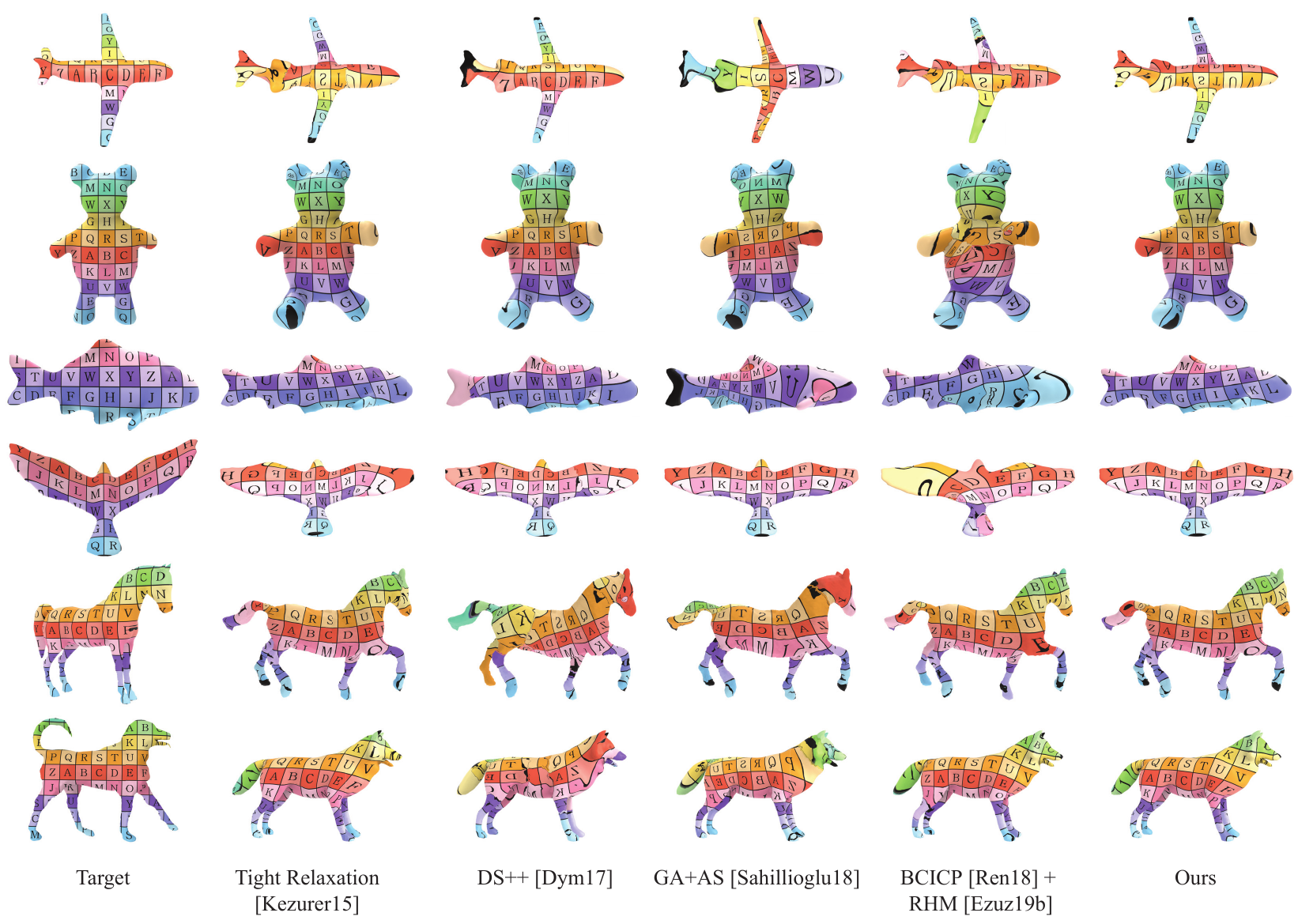}
		\caption{Qualitative comparison between our final results and fully 
			automatic state of the art 
			methods~\cite{kezurer2015tight,DSppDym,SahilliogluGenetic,Ren2018Continuous}
			(after applying the same post processing~\cite{RHM} on all the methods to 
			extract a dense map). The pointwise maps are visualized using texture that 
			is computed on the target mesh (left) and transferred to the source using 
			each method.} 
		\label{fig:shrec07_texture_comp}
	\end{figure*}

	We compare our method with state of the art methods for sparse and dense
	correspondence. We only compare with \emph{fully automatic} methods, since semi automatic methods require additional user input, which is often not available. For comparisons that require a dense correspondence we apply the same post processing for all methods: we compute a functional map as 
	described in section~\ref{sec:fitness} and use it as input to the sparse-to-dense post processing method~\cite{RHM}. We compare the resulting dense pointwise maps quantitatively using the protocol 
	suggested by Kim et al.~\shortcite{kim2011blended} as described in section~\ref{sec:eval_proto}.

	The sparse correspondence methods we compare with are: "Tight Relaxation" by Kezurer et al.~\shortcite{kezurer2015tight}, 
	DS++ by Dym et al.~\shortcite{DSppDym}, and the recent method by Sahillioglu et 
	al.~\shortcite{SahilliogluGenetic} that also uses a genetic algorithm (GA+AS, AS 
	stands for Adaptive Sampling which they use for improved results).
	We additionally compare with the recent dense correspondence method by Ren et 
	al.~\shortcite{Ren2018Continuous} (BCICP). Since Ren et al.~\shortcite{Ren2018Continuous} computes vertex-to-vertex maps, we apply the post processing~\cite{RHM} on their results as well (BCICP+RHM).
	 
	The quantitative results are shown in Figure~\ref{fig:shrec07_graphs}, and it is evident that 
	our method outperforms all previous approaches on this highly challenging non-isometric dataset.
	The sparse correspondence and the functional map are visualized in Figure~\ref{fig:shrec07_color_comp} using the visualization approach described in Section~\ref{sec:eval_proto}. Note that our method consistently yields semantic correspondences on shapes from various classes, even in highly non-isometric cases such as the airplanes, the fish and the dog/wolf pair. This is even more evident when examining the texture transfer results in Figure~\ref{fig:shrec07_texture_comp}, which visualize the dense map. Note the high quality dense map obtained for highly non-isometric meshes such as the fish and the wolf/dog.
	
\begin{figure}[t!]
		\centering
		\includegraphics[width=\linewidth]{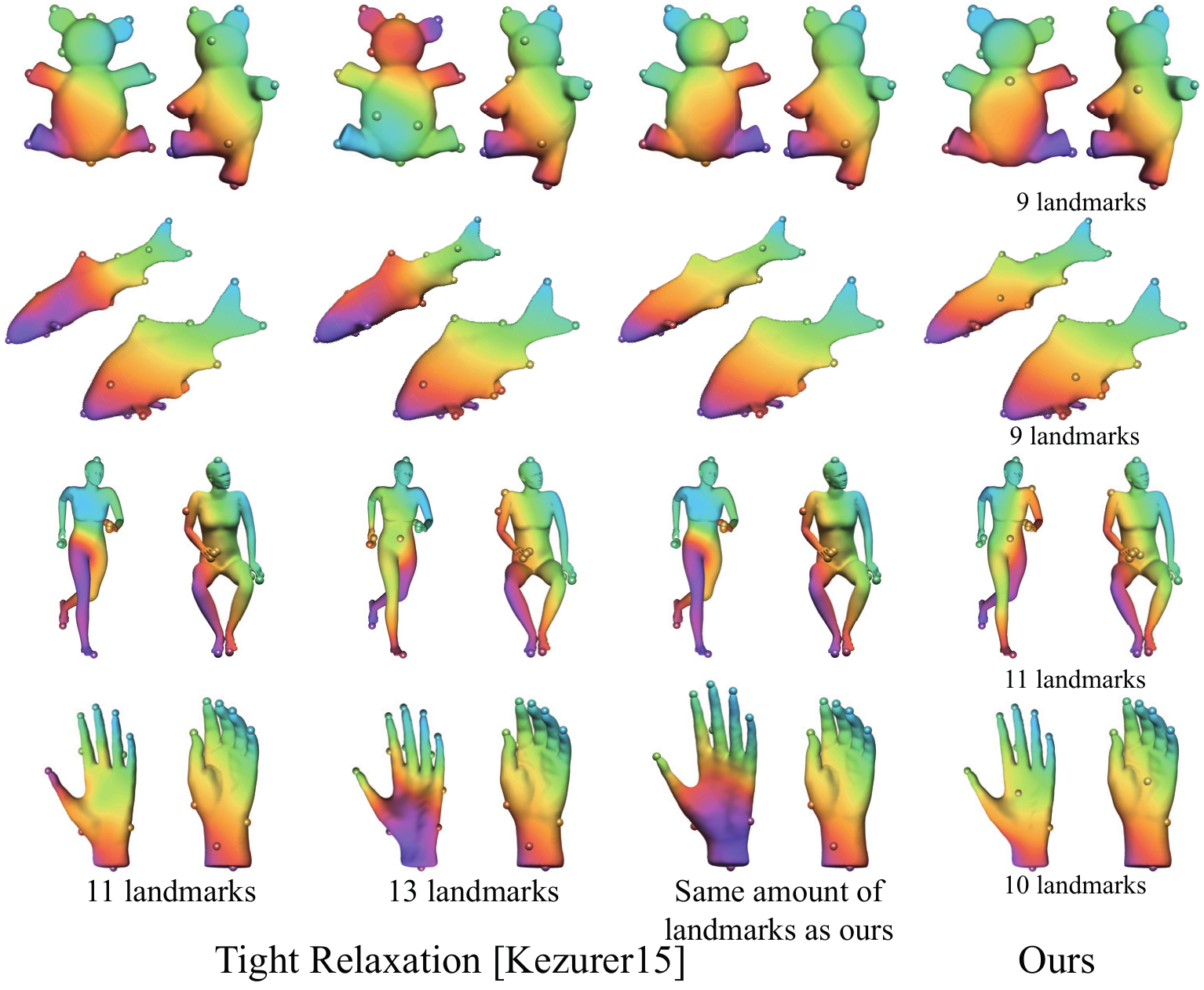}
		\caption{\CH{Qualitative comparison with Tight Relaxation (TR) ~\cite{kezurer2015tight} using the same $15$ landmarks. (right) Our results, (three left columns) TR results with varying match size: $11, 13$ and the size our method found.}} 
		\label{fig:tightLandComp}
	\end{figure}

	\CH{We additionally compare our results to those of "Tight Relaxation" ~\cite{kezurer2015tight} when matching the same initial landmarks. We generate $15$ landmarks for each shape, following the method used in~\cite{kezurer2015tight}. As this approach requires the match size as an additional input, we use the values: $11, 13$ as well as the match size found \emph{automatically} by our method as inputs. The results are shown in Figure~\ref{fig:tightLandComp}. Note that the match found by our method is correct and outperforms the competing output, whose quality also depends on the requested match size.}

	\section{Conclusion}
	We present an approach for \emph{automatically} computing a correspondence between \emph{non-isometric} shapes 
	of the same semantic class. We leverage a genetic 
	algorithm for the combinatorial optimization of a small set of automatically 
	computed landmarks, and use a fitness energy that is based on extending the 
	sparse landmarks to a functional map. As a result, we achieve high quality dense  
	maps, that outperform existing state-of-the-art automatic methods, and can 
	successfully handle challenging cases where the source and target have a different topology.
	
	We believe that our approach can be generalized in a few ways. First, the 
	decomposition of the automatic mapping computation problem into combinatorial 
and continuous problems mirrors other tasks in geometry processing which were 
	handled in a similar manner, such as quadrangular remeshing. It is interesting 
	to investigate whether additional analogues exist between these seemingly 
	unrelated problems. Furthermore, it is intriguing to consider what other 
	problems in shape analysis can benefit from genetic algorithms. One potential 
	example is map synchronization for map collections, where the choice of cycles 
	to synchronize is also combinatorial.

	\section*{Acknowledgments}
	
	The authors acknowledge the support of the German-Israeli Foundation for Scientific Research and Development (grant number I-1339-407.6/2016), the Israel Science Foundation (grant No. 504/16), and the European Research Council (ERC starting grant no. 714776 OPREP). \CH{We also thank SHREC'07, SHREC'19, AIM@SHAPE, Robert W. Sumner and Windows 3D library for providing the models.}

	\bibliographystyle{ACM-Reference-Format}
	\bibliography{ENIGMA}

	\appendix
	\section{Parameters}
	\footnotesize
	\label{sec:Appendix}
	
	All the parameters are unitless and fixed in all the experiments.
 We choose the values experimentally, yet our results are not sensitive to these values. Many of the parameters are required by the genetic algorithm and the functional map computations, which both have standard practices. 
	
	\vspace{-0.1cm}
	\paragraph*{Centers.} To compute the landmarks classified as centers from Eq. ~\eqref{eq:anomalies}, we use $N=30$ eigenfunctions, as suggested by Cheng et al.~\shortcite{cheng2018geometry}.
	\vspace{-0.1cm}
	\paragraph*{Filtering.} We filter close landmarks according to the initial distance of $\lndEps = 0.08$, where the distance is on the normalized shape. In addition, we set the maximal amount of landmarks to $\lndM = 35$. We observed that for a variety of shapes, $35$ landmarks with the specified minimal distance covered the shape features well and gave a large enough subset of matching landmarks for our fitness function.
	\vspace{-0.1cm}
	\paragraph*{Adjacent landmarks.} We use the geodesic distance of $\dAdj = 0.3$ to define adjacent landmarks. This distance is large enough to accommodate highly non-isometric shapes but still allows to construct consistent chromosomes.
	
	\vspace{-0.1cm}
	\paragraph*{Gene bank.} We set the WKS distance to $\epsWks = 0.2$.
	\vspace{-0.1cm}
	\paragraph*{Match size.} The minimal match size is $m_{\min} = \frac{2}{3}m_{\max}$ and the maximal match size is $m_{\max} = \min(\nLands_1, \nLands_2)$ where $\nLands_i$ is the number of landmarks on $M_i$.
	\vspace{-0.1cm}
	\paragraph*{Population construction.} The initial population has $400$ chromosomes, following experiments with varying population sizes and the guidelines by Rylander et al.~\shortcite{rylander2002optimal}.
	\hspace{-0.1cm}
	\vspace{-0.1cm}
	\paragraph*{Functional map optimization.} To construct the functional map we use $k_s = 30, k_t = 60$ and $\alpha = 1$, $\beta = 100$, which is standard practice for functional maps computations.
	\vspace{-0.1cm}
	\paragraph*{Functional map fitness.} For the elastic energy, we set $\gamma = 5\cdot10^{-4}$ similarly to Ezuz et al.~\shortcite{ezuz2019elastic}.
	\vspace{-0.1cm}
	\paragraph*{Genetic operators.} For the operators, we use the rates \pcross$=0.75$, \pmutG$=0.05$ \pmutS$=0.1$, \pmutFMG$=0.05$. The crossover is the main operator, therefore its rate is much higher than the mutations, as is common for genetic algorithms.
	\vspace{-0.1cm}
	\paragraph*{Convergence.} We stop the algorithm if the fittest chromosome remains unchanged for \CH{$50$} iterations or when a maximal iteration number of $700$ iterations is reached.

\end{document}